\documentclass[twocolumn,article,amsmath,amssymb,aps,prd,floatfix,showkeys,superscriptaddress]{revtex4-1}
\usepackage[utf8x]{inputenc} 
\usepackage{bm}
\usepackage{amssymb}
\usepackage{amsmath}
\usepackage{graphicx}
\usepackage{epsfig}
\usepackage{array}
\usepackage{float}
\usepackage{color}
\usepackage[usenames,dvipsnames]{xcolor}
\usepackage{epstopdf}
\usepackage{natbib}
\usepackage{lineno}
\usepackage{slashed}
\usepackage{color}
\usepackage{caption}
\usepackage{hyperref}
\usepackage{slashed}
\usepackage{soul}
\usepackage[bottom]{footmisc}

\newcommand{\be}{\begin{equation}}
\newcommand{\ee}{\end{equation}}
\newcommand{\bea}{\begin{eqnarray}}
\newcommand{\eea}{\end{eqnarray}}
\newcommand{\beas}{\begin{eqnarray*}}
\newcommand{\eeas}{\end{eqnarray*}}
\newcommand{\nn}{\nonumber}
\newcommand{\p}{\parallel}

\newcommand{\gm}{\gamma^{\mu}}
\newcommand{\gn}{\gamma^{\nu}}
\newcommand{\ps}{\slashed{p}}

\newcommand{\ks}{\slashed{k}}

\newcommand{\dpi}{(2\pi)}
\newcommand{\Op}{\mathcal{O}}
\newcommand{\gmn}{g^{\mu\nu}}

\newcommand{\eB}{\left |  q_fB\right |}
\newcommand{\bse}{\begin{subequations}}
\newcommand{\ese}{\end{subequations}}
\newcommand{\kp}{k_\parallel}

\newcommand{\kt}{k_\perp}
\newcommand{\pp}{p_\parallel}
\newcommand{\pt}{p_\perp}

\newcommand{\gmnp}{g^{\mu\nu}_{\parallel}}
\newcommand{\psh}{\slashed{p}}

\newcommand{\Pp}{\mathcal{P}^{\mu\nu}_\parallel}
\newcommand{\Pt}{\mathcal{P}^{\mu\nu}_\perp}
\newcommand{\Pcero}{\mathcal{P}^{\mu\nu}_0}
\newcommand{\factorglobal}{-\frac{i}{4\pi^2}g^2\int d^2x\,}
\newcommand{\pmu}{p^{\mu}}
\newcommand{\pnu}{p^{\nu}}
\newcommand{\Tr}{\text{Tr}}
\newcommand{\kn}{k^\nu}
\newcommand{\km}{k^\mu}

\newcommand{\qt}{q_\perp}

\newcommand{\rhop}{\rho_\parallel}
\newcommand{\rhot}{\rho_\perp}
\newcommand{\B}{\mathcal{B}}

\begin{document}
\title{Gluon polarization tensor in a magnetized medium: Analytic approach starting from the sum over Landau levels}
\author{Alejandro Ayala}
\affiliation{Instituto de Ciencias
  Nucleares, Universidad Nacional Aut\'onoma de M\'exico, Apartado
  Postal 70-543, CdMx 04510,
  Mexico.}
 \affiliation{Centre for Theoretical and Mathematical Physics, and Department of Physics, University of Cape Town, Rondebosch 7700, South Africa.}
\author{Jorge David Casta\~no-Yepes}
\email{Corresponding author.\\
E-mail address: jorgecastanoy@gmail.com (J.D. Casta\~no-Yepes).}
\affiliation{Instituto de Ciencias
  Nucleares, Universidad Nacional Aut\'onoma de M\'exico, Apartado
  Postal 70-543, CdMx 04510,
  Mexico.}
\author{M. Loewe}
\affiliation{Instituto de F\'{\i}sica, Pontificia Universidad Cat\'olica de Chile, Casilla 306, Santiago 22, Chile.}
 \affiliation{Centre for Theoretical and Mathematical Physics, and Department of Physics, University of Cape Town, Rondebosch 7700, South Africa.}
\affiliation{Centro Cient\'\i fico-Tecnol\'ogico de Valpara\'\i so CCTVAL, Universidad T\'ecnica Federico Santa Mar\'\i a, Casilla 110-V, Valapara\'\i so, Chile}
\author{Enrique Mu\~noz}
\affiliation{Instituto de F\'{\i}sica, Pontificia Universidad Cat\'olica de Chile, Casilla 306, Santiago 22, Chile.}
\affiliation{Research Center for Nanotechnology and Advanced Materials CIEN-UC, Pontificia Universidad Cat´olica de Chile, Santiago, Chile.}



\begin{abstract}

We present an analytic method to compute the one-loop magnetic correction to the gluon polarization tensor starting from the Landau-level representation of the quark propagator in the presence of an external magnetic field. We show that the general expression contains the vacuum contribution that can be isolated from the zero-field limit for finite gluon momentum. The general tensor structure for the gluon polarization also contains two spurious terms that do not satisfy the transversality properties. However, we also show that the  coefficients of this structures vanish and thus do not contribute to the polarization tensor, as expected. In order to check the validity of the expressions  we study the strong and weak field limits and show that in the former, the well established result is reproduced. The findings can be used to study the conditions for gluons to equilibrate with the magnetic field produced during the early stages of a relativistic heavy-ion collision.

\end{abstract}

\maketitle

\section{Introduction}\label{I}
The production of hot and dense strongly interacting matter in heavy-ion reactions at high energies, constitutes a driving force for the formulation of novel approaches to study QCD subject to extreme conditions. For semi-central collisions, these conditions include the presence of strong, albeit short-lived, magnetic fields. Many theoretical efforts concentrate on describing these conditions considering that the temperature is the largest of the energy scales~\cite{LQCD1, Nikita, Starinets, Skalozub}. However, it has also been realized that the imprints of these strong fields~\cite{Mclerran, Skokov}, if any, should be searched for studying probes produced during the very early stages of the collision, where the system is not yet equilibrated and the largest of the energy scales is instead the magnetic field itself. Possible imprints include an enhanced prompt photon production and/or the chiral magnetic effect~\cite{McLerran2,ayalacastano,Basar1,Basar2,Zakharov, Tuchin}.

The early stages of a heavy-ion reaction are also characterized by the presence of a large number of low momentum gluons which are thought to give rise to the saturation phenomenon described by the Glasma~\cite{Glasma}. When a magnetic field is present, gluon dynamics can also be affected. A deeper understanding of gluon properties within a magnetized medium is crucial to describe the evolution of observables coming from these early stages.

The gluon dispersive properties in a magnetized medium are encoded in the gluon polarization tensor $\Pi^{\mu\nu}$. In a perturbative approach, deviations from its vacuum properties come from the coupling of the magnetic field to virtual quarks. The quark propagator can be represented in terms of a sum over Landau levels. When the field is strong, calculations often resort to the approximation where these quarks occupy the lowest Landau level (LLL), which simplifies considerably the treatment~\cite{Fukushima, Bandyopadhyay, ADHHLMZ}. Nevertheless, when the field is not as intense, it is important to perform a sum over Landau levels to capture effects that may be missing from expressions restricted to the LLL, in particular, the emergence of tensor polarization structures other than the parallel one that make up the full polarization tensor. This kind of calculations have been performed at one-loop level for the photon polarization tensor~\cite{Hattori} in the context of the vacuum birefringence in strong magnetic fields, where the authors resort to a numerical treatment for the infinite sum over Landau levels. However, in order to gain a deeper insight, an analytical approach for the infinite sum over Landau levels is desirable. In this work, we undertake such task and present an analytic method to perform the sum over all Landau levels for the coefficients of the tensor structure that make up the gluon polarization tensor in the presence of a magnetic field of arbitrary intensity. The vacuum contribution is obtained in the limit when $B\to 0$. We show that by this procedure one obtains the usual fermion contribution to the vacuum polarization tensor, together with a second term that is shown to vanish, given the properties of its coefficient under scaling transformations. Applying the same argument to the full, magnetic field-dependent polarization tensor, it is possible to isolate the physical tensor structures and their coefficients, thus getting rid of spurious terms. We then proceed to carefully subtract the vacuum pieces to remove ultraviolet divergences. The procedure ensures that the remaining, magnetic field dependent contributions are finite. In order to test the validity of the expressions thus obtained, we study the weak and strong magnetic field limits. The work is organized as follows: In Sec.~\ref{II}, we write the one-loop expression for the gluon polarization tensor in the presence of a constant external magnetic field. We chose the tensor basis to express the polarization tensor and outline the calculation to carry out the product of fermion propagators and the corresponding sums over Landau levels. We show that after the sum is made, there appear two spurious, non-transverse terms. These are shown to vanish, as in the vacuum case, from the properties of their coefficients under scaling transformations. In Sec.~\ref{III} we study the strong and in Sec.~\ref{IV} the weak field limits and show that the obtained expressions coincide with well known results. We summarize and discuss our results in Sec.~\ref{V} and leave for the appendices the calculation details. 

\section{Gluon Polarization tensor}\label{II}
\begin{figure}[t]
    \centering
    \includegraphics[scale=0.33]{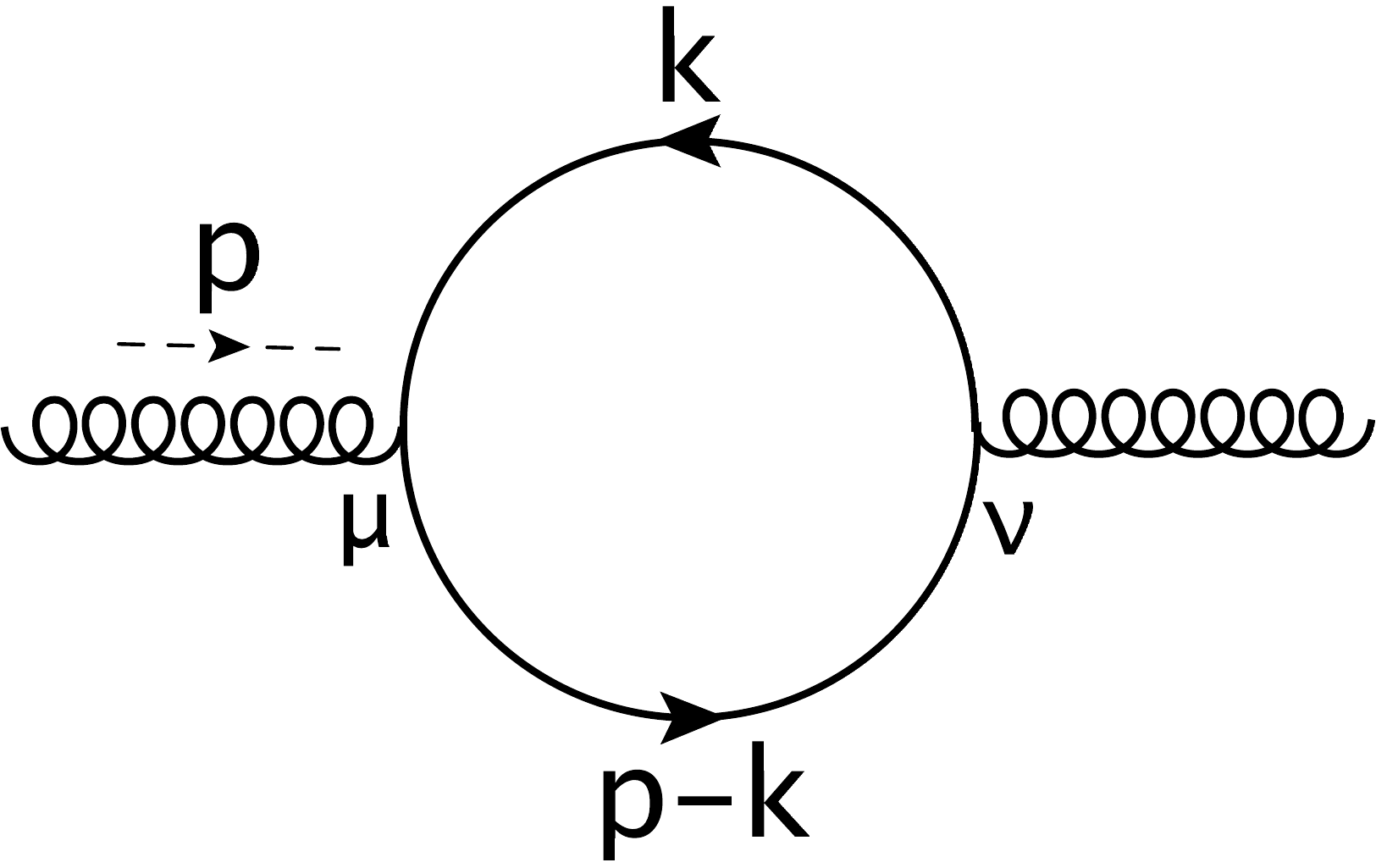}
    \caption{One-loop diagram representing the gluon polarization
tensor.}
    \label{diagrama}
\end{figure}

We start from the one-loop contribution to the gluon polarization tensor, which is depicted in Fig.~\ref{diagrama} and is given explicitly by
\bea
i\Pi^{\mu\nu}_{ab}&=&-\frac{1}{2}\int\frac{d^4k}{\dpi^4}\text{Tr}\left\{igt_b\gamma^\nu iS^{(n)}(k)igt_a\gamma^\mu iS^{(m)}(q)\right\}\nn\\
&+& {\mbox{C.C.}},
\label{Pidef}
\eea
where C.C. refers to the {\it charge conjugate} contribution, that is, the contribution where the flow of charge within the loop is in the opposite direction. The factor $1/2$ accounts for the symmetry factor, which in the presence of the external magnetic field comes about given that the two contributing diagrams, with the opposite flow of charge, are not equivalent. Also $g$ is the strong coupling. $S(k)$ is the quark propagator and $t_{a,b}$ are the generators of the color group in the fundamental representation. The fermion propagator in the presence of a magnetic field $\vec{B}=B\hat{z}$ can be written in terms of a sum over Landau levels as~\cite{MiranskiPropagador1,MiranskiPropagador2}
\bea
iS(p)=i e^{-\pt^2/\eB}\sum_{n=0}^{+\infty}(-1)^n\frac{D_n(q_fB,p)}{p_\p^2-m_f^2-2n\left |  q_fB\right |},
\label{fermionpropdef}
\eea
where
$m_f$ and $q_f$ are the quark mass and electric charge, respectively, and
\bea
D_n(q_fB,p)&=&2(\psh_\p+m_f)\mathcal{O}^{-}L_n^0\left(\frac{2\pt^2}{\left | q_fB \right|}\right)\nn\\
&-&2(\psh_\p+m_f)\mathcal{O}^{+}L_{n-1}^0\left(\frac{2\pt^2}{\left | q_fB \right |}\right)\nn\\
&+&4\psh_\perp L_{n-1}^1\left(\frac{2\pt^2}{\left | q_fB \right |}\right).
\label{Dn}
\eea
In Eq.~(\ref{Dn}), $L_n^\alpha(x)$ are the generalized Laguerre polynomials, with the index $n$ labeling the $n$-th Landau level, and 
\bea
\mathcal{O}^{(\pm)}=\frac{1}{2}\left[1\pm i\gamma^1\gamma^2\text{sign}(q_f B)\right].
\label{Op}
\eea
Also, we follow the convention whereby the square of the four-momentum $p^\mu$, expressed in terms of the square of its parallel and perpendicular (with respect to the magnetic field direction) components, is given by 
\bea
   p^2=p_\parallel^2-\pt^2=(p_0^2-p_3^2)-(p_1^2+p_2^2).
\eea
Computing Eqs. (\ref{Pidef}) and (\ref{fermionpropdef}), after performing the sum over all Landau levels, the gluon polarization tensor can be written in terms of four tensor structures, given by
\bea
\!\!\!\!\!\!i\Pi^{\mu\nu}=
\factorglobal f_0\left(x_1,x_2\right)\sum_{i=1}^4f_i^{\mu\nu}(x_1,x_2),
\label{tensorfirst}
\eea
where on the right-hand side, we have omitted a factor $\delta_{ab}$ coming from using the relation $\Tr (t^at^b)=\delta_{ab}/2$, and correspondingly, for notation simplicity, removed the color indices on the left-hand side. Here $(x_1,x_2)\in(0,\infty)$ are Schwinger parameters, with $d^2x=dx_1dx_2$ and 
\begin{subequations}
\bea
&&\!\!\!\!\!\!\!\!f_0\left(x_1,x_2\right)=\exp\left[\frac{x_1 x_2}{x_1+x_2}\pp^2-m_f^2(x_1+x_2)\right]\nn\\
&\!\!\!\!\!\!\!\!\!\!\!\!\!\!\!\!\!\!\!\!\!\times\!\!\!\!\!\!\!\!\!\!\!\!\!\!&\exp\left[-\frac{\tanh(\eB x_1)\tanh(\eB x_2)}{\tanh(\eB x_1)+\tanh(\eB x_2)}\frac{\pt^2}{\eB}\right],
\label{f0}
\eea

\bea
&&f_1^{\mu\nu}(x_1,x_2)=\eB\coth\left[\eB(x_1+x_2)\right]\nn\\
&\times&\left[\left(\frac{x_1x_2}{(x_1+x_2)^3}\pp^2+\frac{m_f^2}{x_1+x_2}\right)\gmn_\parallel-\frac{2x_1x_2}{(x_1+x_2)^3}\pmu_\parallel\pnu_\parallel\right],\nn\\
\label{f1}
\eea

\bea
&&f_2^{\mu\nu}(x_1,x_2)=\eB\frac{\cosh\left[\eB(x_2-x_1)\right]}{\sinh\left[\eB(x_1+x_2)\right]}\nn\\
&\times&\left[\frac{x_1x_2}{(x_1+x_2)^3}\pp^2+\frac{m_f^2}{x_1+x_2}+\frac{1}{(x_1+x_2)^2}\right]\gmn_\perp,
\label{f2}
\eea

\bea
&&f_3^{\mu\nu}(x_1,x_2)=\frac{\eB}{2(x_1+x_2)^2\sinh^2\left[\eB (x_1+x_2)\right]}\nn\\
&\times&\Big[x_1\sinh(2\eB x_2)+x_2\sinh(2\eB x_1)\Big]\nn\\
&\times&\left(\pmu_\parallel\pnu_\perp+\pnu_\parallel\pmu_\perp\right),
\label{f3}
\eea

\bea
&&f_4^{\mu\nu}(x_1,x_2)=\frac{\eB^2}{(x_1+x_2)\sinh^2\left[\eB(x_1+x_2)\right]}\nn\\
&\times&\left[
\left(1-\frac{\tanh(\eB x_1)\tanh(\eB x_2)}{\eB\left[\tanh(\eB x_1)+\tanh(\eB x_2)\right]}\pt^2\right)\gmn\right.\nn\\
&-&\gmn_\perp-\left.\frac{2\tanh(\eB x_1)\tanh(\eB x_2)}{\eB\left[\tanh(\eB x_1)+\tanh(\eB x_2)\right]}\pmu_\perp\pnu_\perp\right].\nn\\
\label{f4}
\eea
\label{fs}
\end{subequations}
For calculation details, see Appendix~\ref{ApA}.
\subsection{Tensor Basis}
The gluon polarization tensor should be represented by a symmetric tensor under the exchange of its Lorentz indices. It can be constructed out of the external products of the independent vectors describing the propagation of a gluon with momentum $p^\mu$ in the presence of a magnetic field whose direction is specified by a four-vector $b^\mu$, in addition to the metric tensor $g^{\mu\nu}$. Without loss of generality, we can choose a reference frame where the magnetic field points along the $\hat{z}$ axis. Due to the presence of this Lorentz invariance-breaking vector, it is convenient to split the metric itself into parallel and perpendicular (with respect to the magnetic field direction) components, that is
\bea
g^{\mu\nu}=g^{\mu\nu}_\parallel + g^{\mu\nu}_\perp,
\label{metric}
\eea
where 
\bea
g^{\mu\nu}_\parallel={\mbox{diag}}(1,0,0,-1),
\label{metricpara}
\eea
and
\bea
g^{\mu\nu}_\perp={\mbox{diag}}(0,-1,-1,0).
\label{metricperp}
\eea
We thus see that the most general symmetric tensor can be constructed out of combinations of the {\it four} possible independent tensors
\bea
p^\mu p^\nu,\  b^\mu b^\nu,\ p^\mu b^\nu + p^\nu b^\mu,\ g^{\mu\nu}.
\label{possible}
\eea
However, notice that in QCD, $\Pi^{\mu\nu}$ must satisfy the generalized Ward-Takahashi identity namely, the {\it transversality} condition
\bea
p_\mu p_\nu \Pi^{\mu\nu}=0.
\label{WTI}
\eea
Therefore, since Eq.~(\ref{WTI}), implies a relation between the coefficients of the tensors to express $\Pi^{\mu\nu}$, only three {\it transverse} tensors turn out to be independent. To visualize this, let us suppose that $\Pi^{\mu\nu}$ can be written as
\bea
\Pi^{\mu\nu} = a\ A^{\mu\nu} + b\ B^{\mu\nu} + c\ C^{\mu\nu} + d\ D^{\mu\nu}.
\eea
Gauge invariance, Eq.~(\ref{WTI}), implies
\bea
p_{\mu} p_{\nu} \Pi^{\mu\nu} &=& a \left(p_{\mu} p_{\nu} A^{\mu\nu}\right)+ b \left(p_{\mu} p_{\nu} B^{\mu\nu}\right)\nn\\
&+& c \left(p_{\mu}p_{\nu}C^{\mu\nu}\right)+ d \left(p_{\mu} p_{\nu} D^{\mu\nu}\right) = 0,
\label{newimply}
\eea
Equation~(\ref{newimply}) means that only three out of the four factors $(a,b,c,d)$ are independent. Therefore, the tensor structure that multiplies the factor chosen as not independent can be distributed among the rest of the structures to result in only three of them being needed to span the whole tensor $\Pi^{\mu\nu}$. A convenient basis to express the polarization tensor is such that the independent tensors are chosen each to be {\it transverse}, in such a way that Eq.~(\ref{WTI}) be satisfied already as
\bea
p_\mu \Pi^{\mu\nu}=0.
\label{WTI2}
\eea
This choice has the advantage that the basis can be used to express the polarization tensor either in QCD or in QED.
In the present work, we chose the orthonormal basis
\bea
\mathcal{P}_{\p}^{\mu\nu}=\gmn_{\p}-\frac{p_\p^\mu p_\p^\nu}{p_\p^2},
\label{pipara}
\eea
\bea
\mathcal{P}_{\perp}^{\mu\nu}=\gmn_{\perp}+\frac{\pt^\mu \pt^\nu}{\pt^2},
\label{piperp}
\eea

\bea
\mathcal{P}_{0}^{\mu\nu}=\gmn-\frac{p^\mu p^\nu}{p^2}-\mathcal{P}_{\p}^{\mu\nu}-\mathcal{P}_{\perp}^{\mu\nu}.
\label{pizero}
\eea
Such choice comes from the factorization of the metric into transverse and parallel structures induced by the presence of the vector $b^\mu$ representing the direction of the magnetic field. To show this, we can  choose $b^{\mu}=(0,\mathbf{b})=(0,0,0,1)$. Introducing the space-vector $\mathbf{a}=(1/2)(-y,x,0)$ such that $\mathbf{b}=\nabla\times\mathbf{a}$, we observe that by choosing the vector potential as $\mathbf{A}=B\mathbf{a}$, and from the definition $F^{\mu\nu}=\partial^\mu A^\nu -  \partial^\nu A^\mu $ we get
\bea
p_{\alpha} p_{\beta} F^{\alpha\mu} F^{\beta\nu} =B^2\begin{pmatrix}
0 & 0 & 0 &0 \\ 
0 & p_2^2 & -p_1 p_2 &0 \\ 
0 &- p_1 p_2 & p_1^2 &0 \\ 
0 & 0 &0  & 0
\end{pmatrix}.
\eea
Also, from Eq.~(\ref{piperp})
\bea
\pt^2\Pt=\begin{pmatrix}
0 & 0 & 0 &0 \\ 
0 & -p_2^2 & p_1 p_2 &0 \\ 
0 &p_1 p_2 &- p_1^2 &0 \\ 
0 & 0 &0  & 0
\end{pmatrix},
\eea
therefore
\bea
B^2\pt^2\Pt=-p_{\alpha} p_{\beta} F^{\alpha\mu} F^{\beta\nu},
\eea
which shows that the choice of $b^\mu$ impacts directly the factorization of the metric into transverse and a parallel structures.

On the other hand, notice that when Eqs.~(\ref{pipara})-(\ref{pizero}) are chosen as the basis to span $\Pi^{\mu\nu}$, the condition of Eq.~(\ref{WTI2}) does not reduce the number of independent tensor structures from three to two, given that the tensor structures are already transverse.

Therefore, we can use this basis (see also Ref.~\cite{Angel}) to express Eqs. (\ref{fs}) (see Appendix~\ref{ApB}) as
\bea
&&i\Pi^{\mu\nu}=\factorglobal f_0(x_1,x_2)\nn\\
&\times&\Bigg[\Pi_\parallel\left(x_1,x_2\right)\Pp+\Pi_\perp\left(x_1,x_2\right)\Pt+\Pi_0\left(x_1,x_2\right)\Pcero\nn\\
&+&A_1\left(x_1,x_2\right)\gmn_\parallel+A_2\left(x_1,x_2\right)\gmn_\perp\Bigg],
\label{Pienbaseortonormal}
\eea
where 
\bea
\Pi_\parallel&=&\eB\Bigg[\frac{2x_1x_2\coth\left[\eB(x_1+x_2)\right]}{(x_1+x_2)^3}\pp^2\nn\\
&-&\frac{x_1\sinh(2\eB x_2)+x_2\sinh(2\eB x_1)}{2(x_1+x_2)^2\sinh^2\left[\eB (x_1+x_2)\right]}\pt^2\Bigg],\nn\\
\label{Pipara}
\eea

\bea
\Pi_\perp&=&\eB\Bigg[\frac{x_1\sinh(2\eB x_2)+x_2\sinh(2\eB x_1)}{2(x_1+x_2)^2\sinh^2\left[\eB (x_1+x_2)\right]}\pp^2\nn\\
&-&\frac{2\sinh(\eB x_1)\sinh(\eB x_2)}{(x_1+x_2)\sinh^3\left[\eB(x_1+x_2)\right]}\pt^2\Bigg],
\label{Piperp}
\eea

\bea
\Pi_0=\eB\frac{x_1\sinh(2\eB x_2)+x_2\sinh(2\eB x_1)}{2(x_1+x_2)^2\sinh^2\left[\eB (x_1+x_2)\right]}p^2,\nn\\
\label{Picero}
\eea
 
\bea
&&A_1=\eB\Bigg[\frac{x_1\sinh(2\eB x_2)+x_2\sinh(2\eB x_1)}{2(x_1+x_2)^2\sinh^2\left[\eB (x_1+x_2)\right]}\pt^2\nn\\
&+&\frac{\coth\left[\eB(x_1+x_2)\right]}{(x_1+x_2)^3}\left(m_f^2(x_1+x_2)^2-x_1x_2\pp^2\right)\nn\\
&+&\frac{\eB}{(x_1+x_2)\sinh^2\left[\eB(x_1+x_2)\right]}\nn\\
&\times&\left(1-\frac{\tanh(\eB x_1)\tanh(\eB x_2)}{\eB\left[\tanh(\eB x_1)+\tanh(\eB x_2)\right]}\pt^2\right)\Bigg],\nn\\
\label{coefA}
\eea
and

\bea
A_2&=&\eB\Bigg[\frac{\cosh\left[\eB(x_2-x_1)\right]}{(x_1+x_2)^3\sinh\left[\eB(x_1+x_2)\right]}\nn\\
&\times&\left[x_1x_2\pp^2+(x_1+x_2)+m_f^2(x_1+x_2)^2\right]\nn\\
&-&\frac{x_1\sinh(2\eB x_2)+x_2\sinh(2\eB x_1)}{2(x_1+x_2)^2\sinh^2\left[\eB (x_1+x_2)\right]}\pp^2\nn\\
&+&\frac{\sinh(\eB x_1)\sinh(\eB x_2)}{(x_1+x_2)\sinh^3\left[\eB(x_1+x_2)\right]}\pt^2\Bigg].
\label{coefB}
\eea
Notice that, contrary to expectations, Eq.~(\ref{Pienbaseortonormal}) contains also terms proportional to the tensors $g^{\mu\nu}_\parallel$ and $g^{\mu\nu}_\perp$. In order to show that $\Pi^{\mu\nu}$ is made out only of combinations of transverse tensors, we need to prove that the coefficients $A_1$ and $A_2$ vanish. This is shown in Appendix~\ref{ApC}. For the time being, let us only emphasize that, had we simply projected out Eq.~(\ref{tensorfirst}) onto the basis given by Eqs.~(\ref{Pipara})--(\ref{Picero}), the spurious terms would have induced non-physical contributions that, given their complexity, could obscure the numerical evaluation of the physical coefficients~\cite{Hattori,Hattori2,Ishikawa}. This comes about since, formally, a simple projection would give rise to the tensor coefficients
\begin{subequations}
\bea
\widetilde{\Pi}_\parallel=\Pi_{\mu\nu}\Pp=\Pi_\parallel+A_1,
\label{PiparaconA1}
\eea

\bea
\widetilde{\Pi}_\perp=\Pi_{\mu\nu}\Pt=\Pi_\perp+A_2,
\label{PiperpconA2}
\eea
and
\bea
\widetilde{\Pi}_0=\Pi_{\mu\nu}\Pcero=\Pi_0-\frac{\pt^2}{p^2}A_1+\frac{\pp^2}{p^2}A_2,
\label{PiceroconA1A2}
\eea
where $\Pi_\parallel,\Pi_\perp$ and $\Pi_0$ are given by Eqs.~(\ref{Pipara})-(\ref{Picero}) and $A_1,A_2$ are given by Eqs.~(\ref{coefA})-(\ref{coefB}),
\end{subequations}
showing that such projection contains spurious terms.

\subsection{Vacuum Polarization Tensor}
As one can expect, the gluon polarization tensor contains divergences which come from the vacuum contribution. In order to proceed to isolate these contributions we notice that two possible vacua can be defined:

\begin{itemize}
    \item A vacuum where $p^\mu=0$ and $B=0$, corresponding to a situation where particles and magnetic field appear simultaneously. 
    
    \item A vacuum with $B=0$ and $p^\mu\neq0$, representing a situation where the external field is turned on with pre-existing gluons with four-momentum $p^\mu$.
\end{itemize}

The first choice is ambiguous, given that the energy scales associated to the magnetic field and the transverse momentum appear within the combination $\pt^2/\eB$, and thus, $\pt^2$ and $B$ cannot be set to zero simultaneously. Therefore, we chose to extract the vacuum working in the situation described by the second case. The vacuum contribution is thus given by
\bea
&&i\Pi^{\mu\nu}(p,\eB\rightarrow0)\nn\\
&=&-\frac{i}{8\pi^2}g^2\int d^2x\exp\left[\frac{x_1x_2}{x_1+x_2}p^2-m_f^2(x_1+x_2)\right]\nn\\
&\times&\left[\frac{2x_1x_2}{(x_1+x_2)^4}p^2\left(\gmn-\frac{\pmu\pnu}{p^2}\right)\right.\nn\\
&+&\left.\frac{1}{(x_1+x_2)^3}\left((x_1+x_2)m_f^2-\frac{x_1x_2}{x_1+x_2}p^2+1\right)g^{\mu\nu}\right].\nn\\
\label{vacuum}
\eea
Notice that Eq.~(\ref{vacuum}) contains a term that does not simply vanish under contraction with $p_\mu$, namely, the term proportional to $g^{\mu\nu}$. In order to show that the coefficient of this term vanishes, we follow the argument in Ref.~\cite{Bjorken}. We introduce the scaling transformation for the Schwinger parameters in such a way that $x_i\rightarrow\lambda z_i$, where $\lambda$ is a real parameter. Under this transformation, the coefficient of the term proportional to $g^{\mu\nu}$ becomes
\bea
\mathcal{I}&=&\lambda^2\int\frac{d^2z}{\lambda^2(z_1+z_2)^3}\left(m^2(z_1+z_2)-\frac{z_1z_2}{z_1+z_2}p^2+\frac{1}{\lambda}\right)\nn\\
&\times&\exp\left[\lambda\left(\frac{z_1z_2}{z_1+z_2}p^2-m_f^2(z_1+z_2)\right)\right].
\eea
It is easy to show that the integral $\mathcal{I}$ can also be written as
\bea
\mathcal{I}&=&-\lambda\frac{\partial}{\partial\lambda}\int\frac{d^2z}{\lambda(z_1+z_2)^3}e^{\lambda\left(\frac{z_1z_2}{z_1+z_2}p^2-m_f^2(z_1+z_2)\right)}.
\eea
If we now scale back $z_i\to x_i/\lambda$ we observe that the integral becomes $\lambda$-independent and thus its derivative with respect to $\lambda$ vanishes, namely
\bea
\mathcal{I}&=&-\lambda\frac{\partial}{\partial\lambda}\int \frac{d^2x}{(x_1+x_2)^3}e^{\frac{x_1x_2}{x_1+x_2}p^2-m_f^2(x_1+x_2)}\nn\\
&=&0.
\eea
Therefore, the vacuum polarization tensor becomes
\bea
&&i\Pi^{\mu\nu}(p,\eB\rightarrow0)\nn\\
&=&-\frac{i}{8\pi^2}g^2\int d^2x\exp\left[\frac{x_1x_2}{x_1+x_2}p^2-m_f^2(x_1+x_2)\right]\nn\\
&\times&\frac{2x_1x_2}{(x_1+x_2)^4}p^2\left(\gmn-\frac{\pmu\pnu}{p^2}\right)
\label{vacuumtrue1}
\eea
Notice that Eq.~(\ref{vacuumtrue1}) can also be written as
\bea
&&i\Pi^{\mu\nu}(p,\eB\rightarrow0)\nn\\
&=&-\frac{i}{8\pi^2}g^2\int d^2x\exp\left[\frac{x_1x_2}{x_1+x_2}p^2-m_f^2(x_1+x_2)\right]\nn\\
&\times&\frac{2x_1x_2}{(x_1+x_2)^4}p^2\left(\Pcero+\Pp+\Pt\right),
\label{vacuumtrue2}
\eea
where $\Pcero$, $\Pp$ and $\Pt$ are given by Eqs.~(\ref{pipara})--(\ref{pizero}). 

A similar argument is valid for a non-vanishing magnetic field. This means that the coefficients $A_1$ and $A_2$, in Eqs.~(\ref{coefA}) and (\ref{coefB}), respectively, do not contribute to $\Pi^{\mu\nu}$, since they vanish. The systematic evaluation of these terms is shown in Appendix~\ref{ApC}. Thus, the full polarization tensor with the desired physical properties is given by
\bea
&&i\Pi^{\mu\nu}=\factorglobal f_0(x_1,x_2)\nn\\
&\times&\!\!\Bigg[\Pi_\parallel\left(x_1,x_2\right)\Pp\!\!+\Pi_\perp\left(x_1,x_2\right)\Pt\!\!+\Pi_0\left(x_1,x_2\right)\Pcero\Bigg],\nn\\
\label{PisinAyB}
\eea
where $\Pi_\parallel$, $\Pi_\perp$ and $\Pi_0$ are given by Eqs.~(\ref{Pipara}),~(\ref{Piperp}) and~(\ref{Picero}), respectively.

To cancel the vacuum piece, we  subtract from Eq.~(\ref{PisinAyB}) the contribution from Eq.~(\ref{vacuumtrue1}). Therefore, the finite, magnetic field-dependent part of the gluon polarization tensor is explicitly given by
\begin{widetext}

\bea
&&i\Pi^{\mu\nu}=-\frac{i\eB}{8\pi^2}g^2\int\frac{d^2x}{(x_1+x_2)^2} \exp\left[\frac{x_1 x_2}{x_1+x_2}\pp^2-m_f^2(x_1+x_2)\right]\exp\left[-\frac{\tanh(\eB x_1)\tanh(\eB x_2)}{\tanh(\eB x_1)+\tanh(\eB x_2)}\frac{\pt^2}{\eB}\right]\nn\\
&\times&\Bigg{\{}\left[\frac{2x_1x_2\coth\left[\eB(x_1+x_2)\right]}{(x_1+x_2)}\pp^2-\frac{x_1\sinh(2\eB x_2)}{\sinh^2\left[\eB (x_1+x_2)\right]}\pt^2-\widetilde{\Pi}(x_1,x_2)\right]\Pp\nn\\
&+&\left[\frac{x_1\sinh(2\eB x_2)}{\sinh^2\left[\eB (x_1+x_2)\right]}\pp^2-\frac{2(x_1+x_2)\sinh(\eB x_1)\sinh(\eB x_2)}{\sinh^3\left[\eB(x_1+x_2)\right]}\pt^2-\widetilde{\Pi}(x_1,x_2)\right]\Pt\nn\\
&+&\left[\frac{x_1\sinh(2\eB x_2)}{\sinh^2\left[\eB (x_1+x_2)\right]}p^2-\widetilde{\Pi}(x_1,x_2)\right]\Pcero\Bigg{\}},\label{Pifull1}
\eea
where
\bea
\widetilde{\Pi}(x_1,x_2)=\frac{2p^2}{\eB}\frac{x_1x_2}{(x_1+x_2)^2}\exp\left(\frac{\tanh(\eB x_1)\tanh(\eB x_2)}{\tanh(\eB x_1)+\tanh(\eB x_2)}\frac{\pt^2}{\eB}-\frac{x_1x_2}{x_1+x_2}\pt^2\right),
\label{Pifull}
\eea
\end{widetext}
and we have used the symmetry of the integral under the exchange $x_1\leftrightarrow x_2$. In order to check the validity of the above expression, we  proceed to study its limits in the strong and weak magnetic field cases.

\begin{widetext}
\section{Strong field limit}\label{III}
In order to study the strong field limit, let us first introduce the dimensionless variables
\begin{subequations}
\bea
y_i\equiv m_f^2 x_i,\hspace{0.2cm}\rho^2_{\parallel,\perp}\equiv\frac{p_{\parallel,\perp}^2}{m_f^2},\hspace{0.2cm}\mathcal{B}\equiv\frac{\eB}{m_f^2}
\eea
and the new variables $s$ and $y$ related to $y_1$ and $y_2$ by
\bea
y_1\equiv s(1-y),\hspace{0.3cm}y_2\equiv s y,
\eea
\label{variablesvys}
\end{subequations}
so that Eq.~(\ref{PisinAyB}) becomes

\bea
i\Pi^{\mu\nu}&=&-\frac{ig^2m_f^2}{8\pi^2}\int_0^1dy\int_0^{\infty}\,ds\exp\left[s\left(y(1-y)\rhop^2-1\right)\right]\exp\left[-\frac{\cosh(\B s)-\cosh\left[\B s (2y-1)\right]}{2\sinh(\B s)}\frac{\rhot^2}{\B}\right]\nn\\
&\times&\Bigg{\{}\B\left[2y(1-y)\coth(\B s)\rhop^2-\frac{(1-y)\sinh(2\B sy)}{\sinh^2(\B s)}\rhot^2\right]\Pp\nn\\
&+&\B\left[\frac{(1-y)\sinh(2\B sy)}{\sinh^2(\B s)}\rhop^2-\frac{\cosh(\B s)-\cosh\left[\B s (2y-1)\right]}{\sinh^3(\B s)}\rhot^2\right]\Pt+\frac{(1-y)\B\sinh(2\B sy)}{\sinh^2(\B s)}\rho^2\Pcero\Bigg{\}}.
\label{PienSyY}
\eea
\end{widetext}
Note that in the strong field limit
\bea
\B\coth(\B s)\sim\B,
\nonumber\\
\frac{\B\sinh\left(2\B sy\right)}{2\sinh^2(\B s)}\sim0,
\nonumber\\
\frac{\cosh(\B s)-\cosh\left[\B s (2y-1)\right]}{2\sinh(\B s)}\sim\frac{1}{2\B}
\eea
which hold for all $s$ and 
$0<y<1$.
Therefore 
\bea
&&i\Pi^{\mu\nu}=-\frac{ig^2m_f^2\B\rhop^2}{4\pi^2}e^{-\rhot^2/2\B}\nn\\
&\times&\int_{0}^{1}dy\;y(1-y)\int_0^{\infty}\,ds\exp\left[s\left(y(1-y)\rhop^2-1\right)\right]\Pp.\nn\\
\label{Pistrong}
\eea


For the kinematical region such that $y(1-y)\rhop^2<1$, the integration over $s$ can be performed, yielding
\bea
\!\!\!\!\!\!\!\!\!\!i\Pi^{\mu\nu}&=&\frac{ig^2m_f^2\B}{4\pi^2}e^{-\rhot^2/2\B}\int_0^1dy\frac{y(1-y)}{y(1-y)-\rho_\parallel^{-2}}\Pp\nonumber\\
&\equiv&\frac{ig^2m_f^2\B}{4\pi^2}e^{-\rhot^2/2\B}I(\rho_\parallel^2)\Pp
\label{resultadoFukushima}
\eea
which coincides with the result obtained in Refs.~\cite{Fukushima, Bandyopadhyay, ADHHLMZ} where the gluon polarization tensor is computed by considering only the contribution from the LLL.

\begin{figure}[t!]
    \centering
    \includegraphics[scale=0.413]{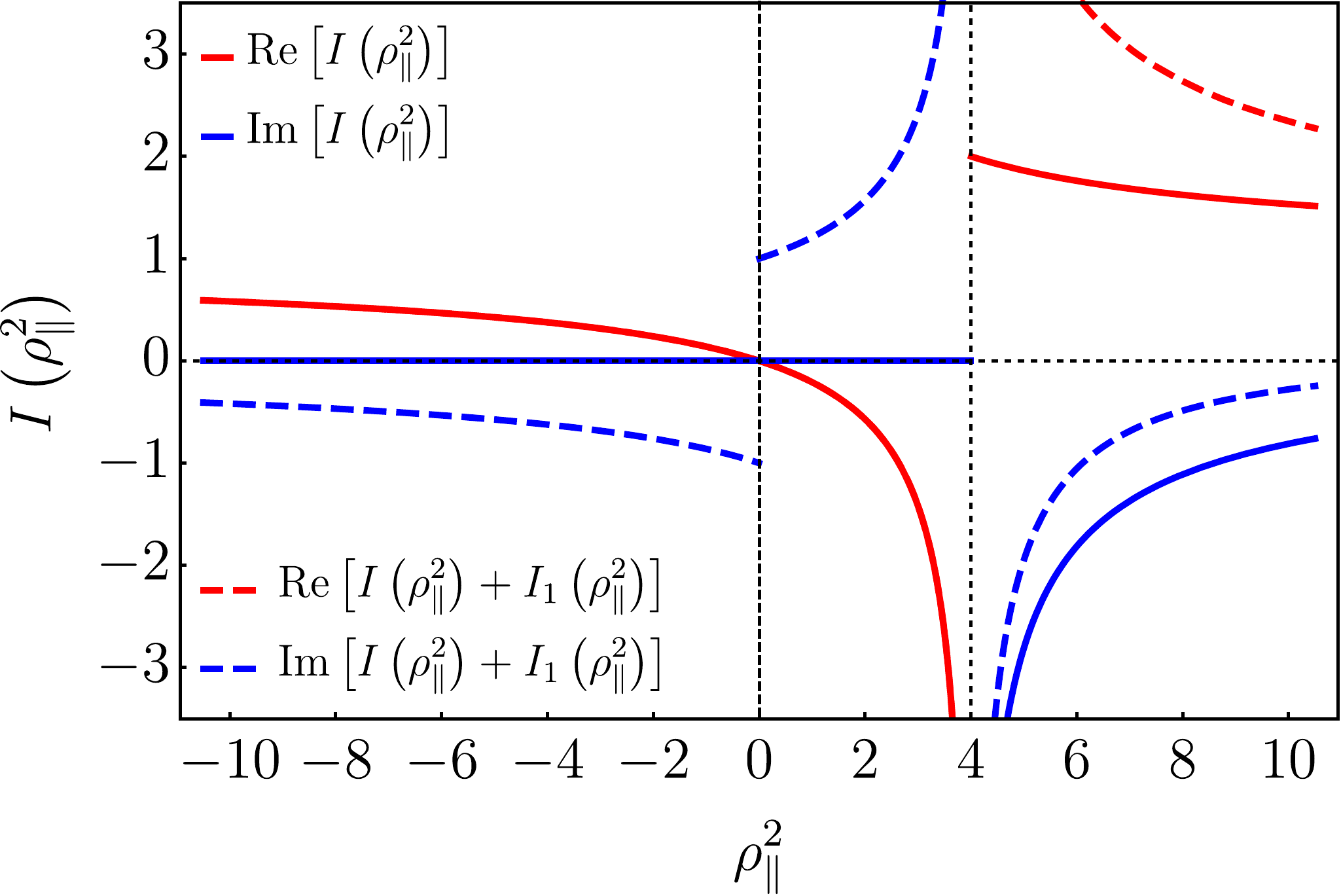}
    \caption{Real and imaginary parts of the function $I(\rho_\parallel^2)$ defined in Eq. (\ref{resultadoFukushima}). Notice the discontinuity at the threshold value $\rho_\parallel=4$ or equivalently at $p_\parallel^2=4m_f^2$. The result including the spurious contribution of $I_1(\rho_\parallel^2)$ from Eq.~(\ref{StrongFieldwithA1}) is also plotted for comparison.}
    \label{Fig:Strong_Field_Limit}
\end{figure}

Figure~\ref{Fig:Strong_Field_Limit} shows the real and imaginary parts of $I(\rho_\parallel^2)$ compared with the result obtained when the contribution of $A_1$, according to Eq.~(\ref{coefA}) is considered. From Eq.~(\ref{PiparaconA1}) the spurious term contributes with
\bea
i\Pi^{\mu\nu}&=&\frac{ig^2m_f^2\B}{4\pi^2}e^{-\rhot^2/2\B}\left[I(\rho_\parallel^2)+I_1(\rho_\parallel^2)\right]\Pp,
\label{StrongFieldwithA1}
\eea
where
\bea
I_1(x)&=&\int_0^1dy\frac{1}{y(1-y)x-1}\nn\\
&=&-\frac{4}{\sqrt{x(x-4)}}\arctan\left(\frac{\sqrt{x}}{\sqrt{4-x}}\right).
\eea

From Fig.~\ref{Fig:Strong_Field_Limit} a discontinuity at the threshold value $\rho_\parallel^2=4$ or equivalently at $p_\parallel^2=4m_f^2$ can be identified. As the figure indicates, the spurious contribution generates an unphysical threshold at $\rhop^2=0$, which cannot be identified with a fermion-pair creation. Notice also that  Eq.~(\ref{Pifull1}) implies the existence of an infinite sequence of momentum thresholds when the external gluon momentum becomes resonant with twice the quark/antiquark {\it magnetic mass}, whose square is defined as $m_{(B)f}^2=m_f^2+2n\eB$. The threshold corresponds to the value of the longitudinal momentum squared for the creation of a quark-antiquark pair, each particle having a magnetic mass corresponding to the given Landau level. 

These thresholds can be obtained from our calculation by concentrating on the conditions where the hyperbolic functions become divergent. For these purposes let us examine the term proportional to $\coth(\B s)$ in Eq.~(\ref{PienSyY})
\bea
\mathcal{K}&=&\int_0^1dy\int_0^{\infty}\,ds\,y(1-y)\B\coth(\B s)\nn\\
&\times&\exp\left[s\left(y(1-y)\rhop^2-1\right)\right]\nn\\
&\times&\exp\left[-\frac{\cosh(\B s)-\cosh\left[\B s (2y-1)\right]}{2\sinh(\B s)}\frac{\rhot^2}{\B}\right].\nn\\
\label{K}
\eea
Notice that if $\B\gg 1$
\bea
&&\B\coth(\B s)\exp\left[-\frac{\cosh(\B s)-\cosh\left[\B s (2y-1)\right]}{2\sinh(\B s)}\frac{\rhot^2}{\B}\right]\nn\\
&=&\B\frac{1+e^{-2\B s}}{1-e^{-2\B s}}\nonumber\\
&\times&\exp\Bigg\{\frac{-1-e^{-2\B s}+e^{-2\B s(y-1)}+e^{-2\B sy}}{2\B\left(1-e^{-2\B s}\right)}\rhot^2\Bigg\}\nn\\
&\approx&\B\frac{1+e^{-2\B s}}{1-e^{-2\B s}}+\mathcal{O}(\rhot^2).
\eea

\begin{figure}[h!]
    \centering
    \includegraphics[scale=0.41]{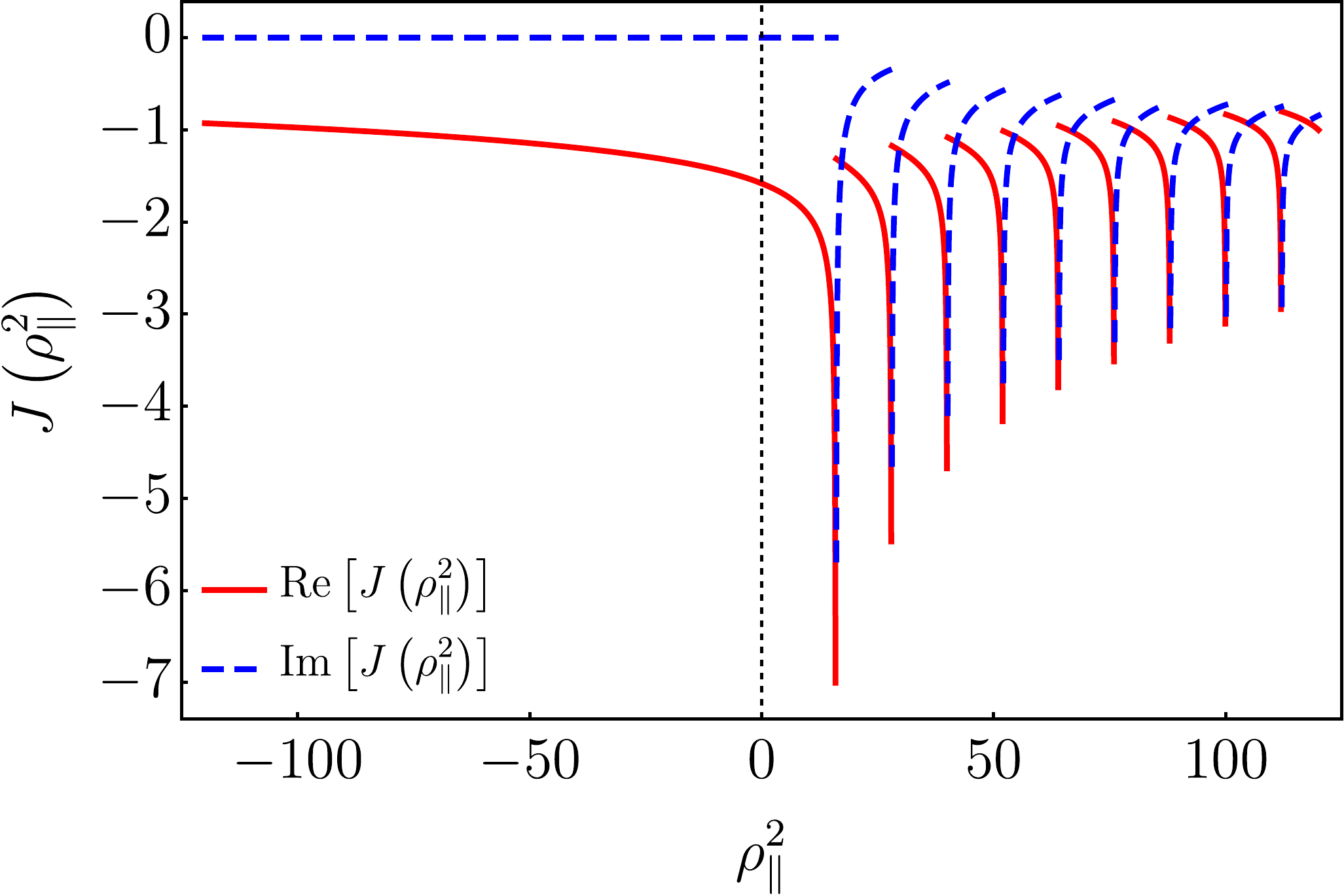}
    \caption{Real and imaginary parts of the function $J(\rhop^2)$ defined in Eq.~(\ref{J(x)}) for $\B=1.5$ and added up to $n_{\text{max}}=100$. Notice the emergence of different resonant thresholds when the quark magnetic mass includes consecutive Landau levels.}
    \label{Fig:UmbralesCoth}
\end{figure}

Using that
\bea
\frac{1}{1-e^{-2\B s}}=\sum_{n=0}^{\infty}e^{-2n\B s},
\label{dominant}
\eea
we can write
\bea
\frac{1+e^{-2\B s}}{1-e^{-2\B s}}=1+2\sum_{n=1}^{\infty}e^{-2n\B s},
\eea
so that, the dominant term in Eq.~(\ref{K}) is given by
\bea
\mathcal{K}&=&\B\int_0^1dy\int_0^{\infty}\,ds\,y(1-y)\Bigg\{\exp\left[s\left(y(1-y)\rhop^2-1\right)\right]\nn\\
&+&2\sum_{n=1}^{\infty}\exp\left[s\left(y(1-y)\rhop^2-2n\B-1\right)\right]\Bigg\}\nn\\
&=&\frac{\B}{\rhop^2}\,I(\rhop^2)+8\B\, J(\rhop^2),
\eea
where $I(x)$ is defined in Eq.~(\ref{resultadoFukushima}) and
\bea
J(x)\equiv-\sum_{n=1}^{\infty}\frac{\arctan\left(\frac{\sqrt{x}}{\sqrt{4(2n\B+1)-x}}\right)}{\sqrt{x\left[4(2n\B+1)-x\right]}}.
\label{J(x)}
\eea
In this way, the resonant behavior of the thresholds is explicit: the gluon polarization tensor has divergences when its momentum reaches the value $\pp^2=4m_{(Bn)f}^2$, where $n$ labels each of the Landau levels. In other words, the creation of quark-antiquark pairs is allowed when the gluon momentum is large enough to generate not only the inertial mass of the pair but rather the magnetic mass, induced by the magnetized medium. Figure~\ref{Fig:UmbralesCoth} shows several thresholds of the function $J(\rhop^2)$ in a broad range of $\rhop^2$ for a maximum value of $n$,  $n_{\text{max}}=100$. The same argument is valid for all terms in Eq.~(\ref{PienSyY}) given that its dominant contribution is given by a power of the series in Eq.~(\ref{dominant}).

\section{Weak field limit}\label{IV}

Let us study the case where the field satisfies the hierarchy of energy scales $|eB| < m_f^2$. We call this the {\it weak field limit}. For this purpose, we can perform a
power series of Eq.~(\ref{PienSyY}) around $\B=0$ to obtain
\begin{widetext}
\bea
i\Pi^{\mu\nu}&=&-\frac{ig^2m_f^2}{8\pi^2}\int_0^1dy\int_0^\infty ds\Bigg{\{}\left[\frac{2y(1-y)}{s}\rho^2+\frac{2sy(1-y)}{3}\left[sy^2(1-y)^2\rhot^2\rho^2+2(1-y^2)\rhot^2+\rho^2\right]\B^2\right]\Pp\nn\\
&+&\left[\frac{2y(1-y)}{s}\rho^2+\frac{2sy(1-y)}{3}\left[sy^2(1-y)^2\rhot^2\rho^2+(1+y)\rhot^2-(1-y^2)\rhop^2+y^2\rho^2\right]\B^2\right]\Pt\nn\\
&+&\left[\frac{2y(1-y)}{s}\rho^2+\frac{2sy(1-y)}{3}\rho^2\left[sy^2(1-y)^2\rhot^2+2y^2-1\right]\B^2\right]\Pcero\Bigg{\}}\exp\left[s\left(y(1-y)\rho^2-1\right)\right],
\label{weakBlimit}
\eea
\end{widetext}
where the vacuum contribution of Eq.~(\ref{vacuumtrue2}) can be identified as
\bea
&&i\Pi^{\mu\nu}\left(\rho^2,\B\rightarrow0\right)\nn\\
&=&-\frac{ig^2m_f^2}{8\pi^2}\int_0^1dy\int_0^\infty ds\exp\left[s\left(y(1-y)\rho^2-1\right)\right]\nn\\
&\times&\frac{2y(1-y)}{s}\rho^2\left(\Pp+\Pt+\Pcero\right)
\eea
\begin{figure*}[t]
    \centering
    \includegraphics[scale=0.4]{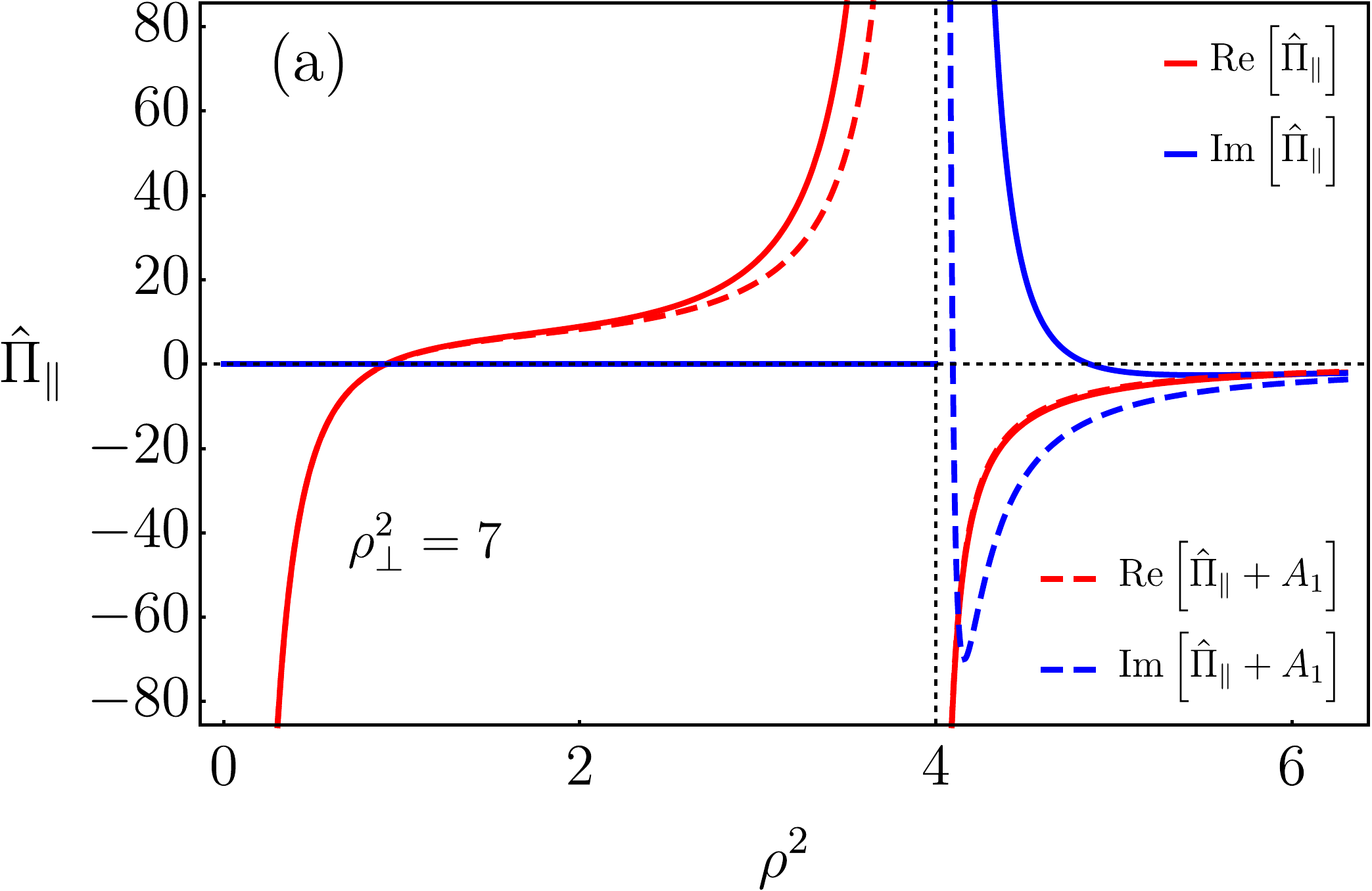}\hspace{0.6cm}\includegraphics[scale=0.4]{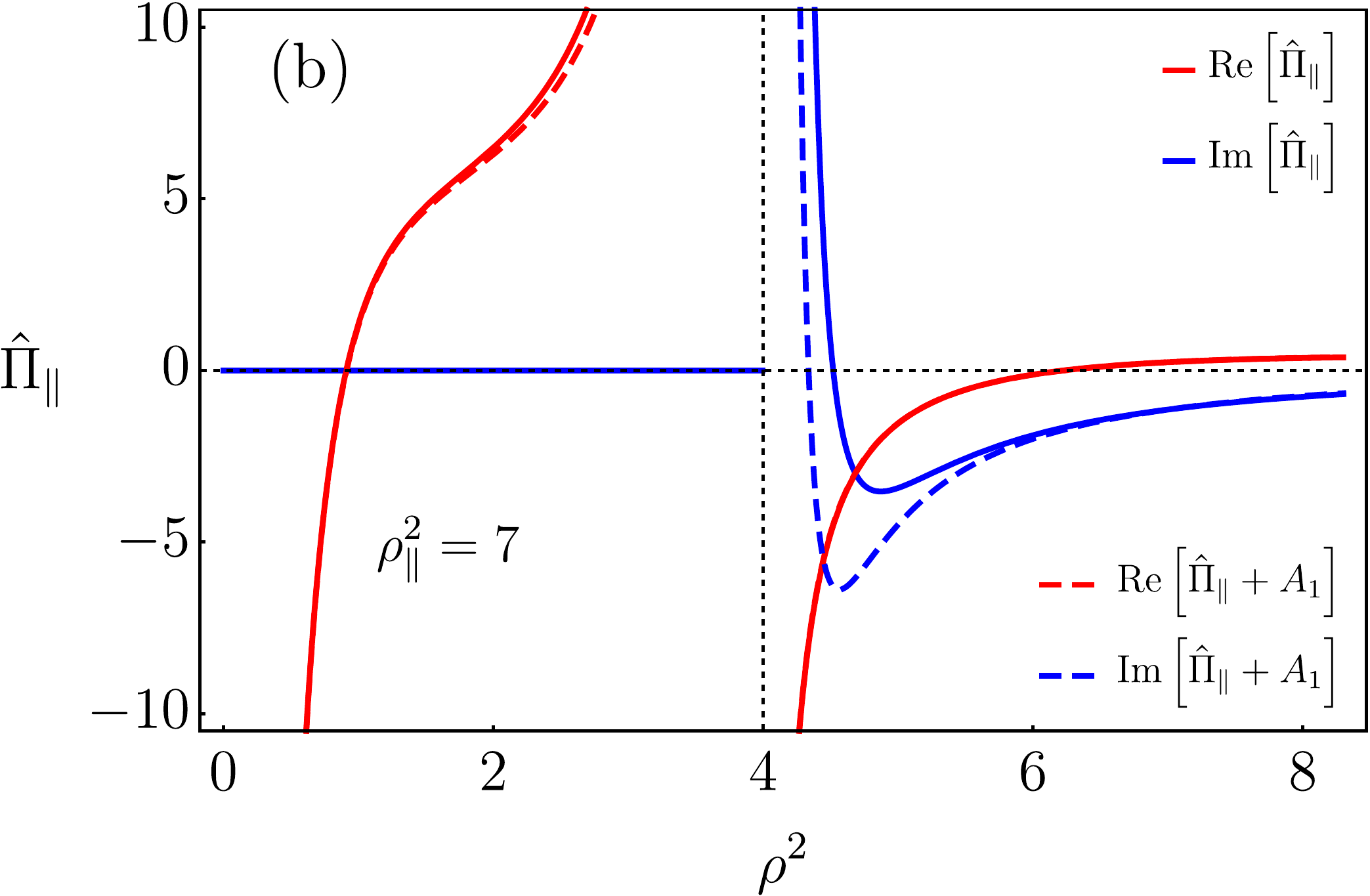}\\
    \vspace{0.5cm}
    \includegraphics[scale=0.4]{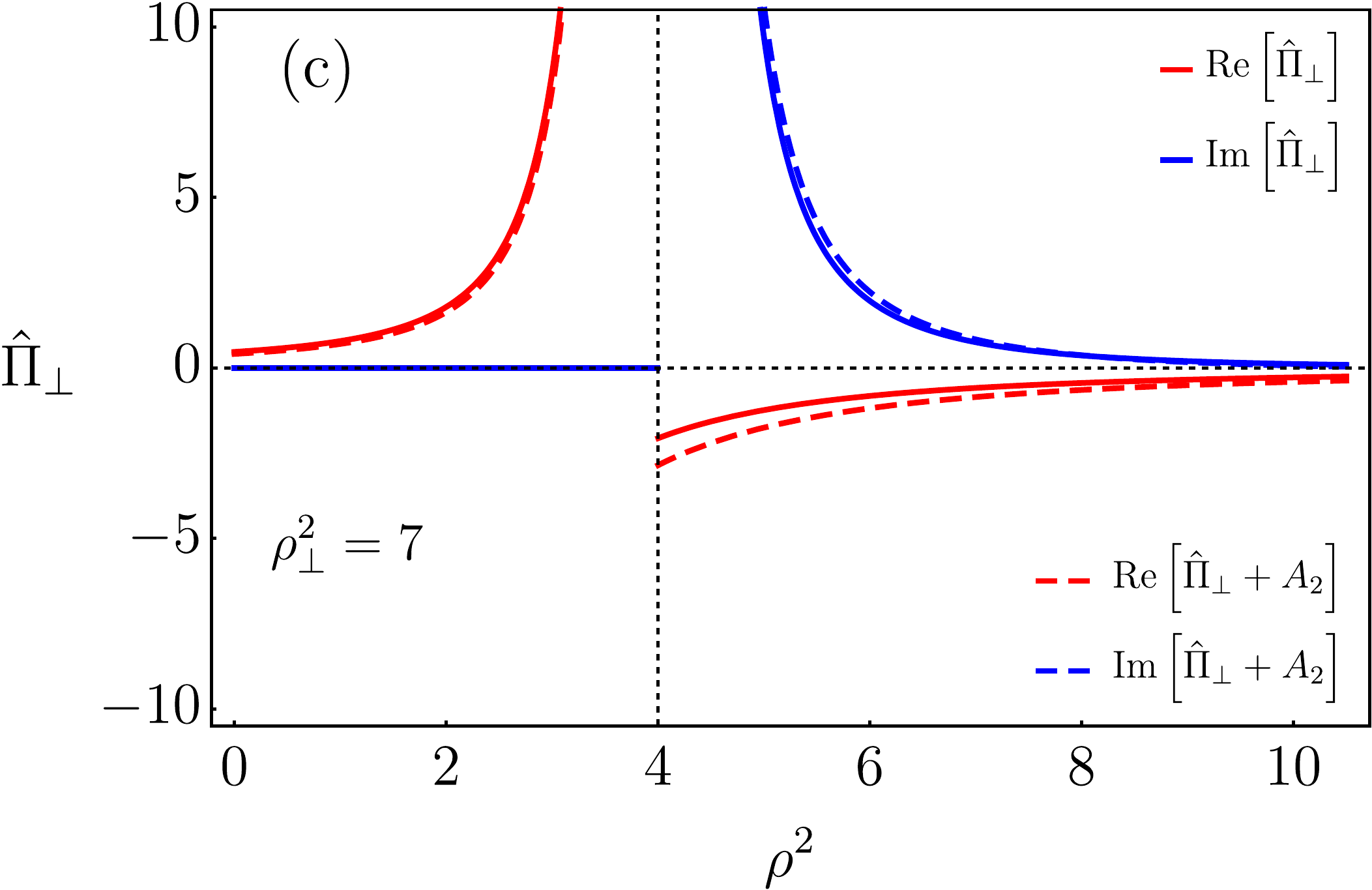}\hspace{0.6cm}\includegraphics[scale=0.4]{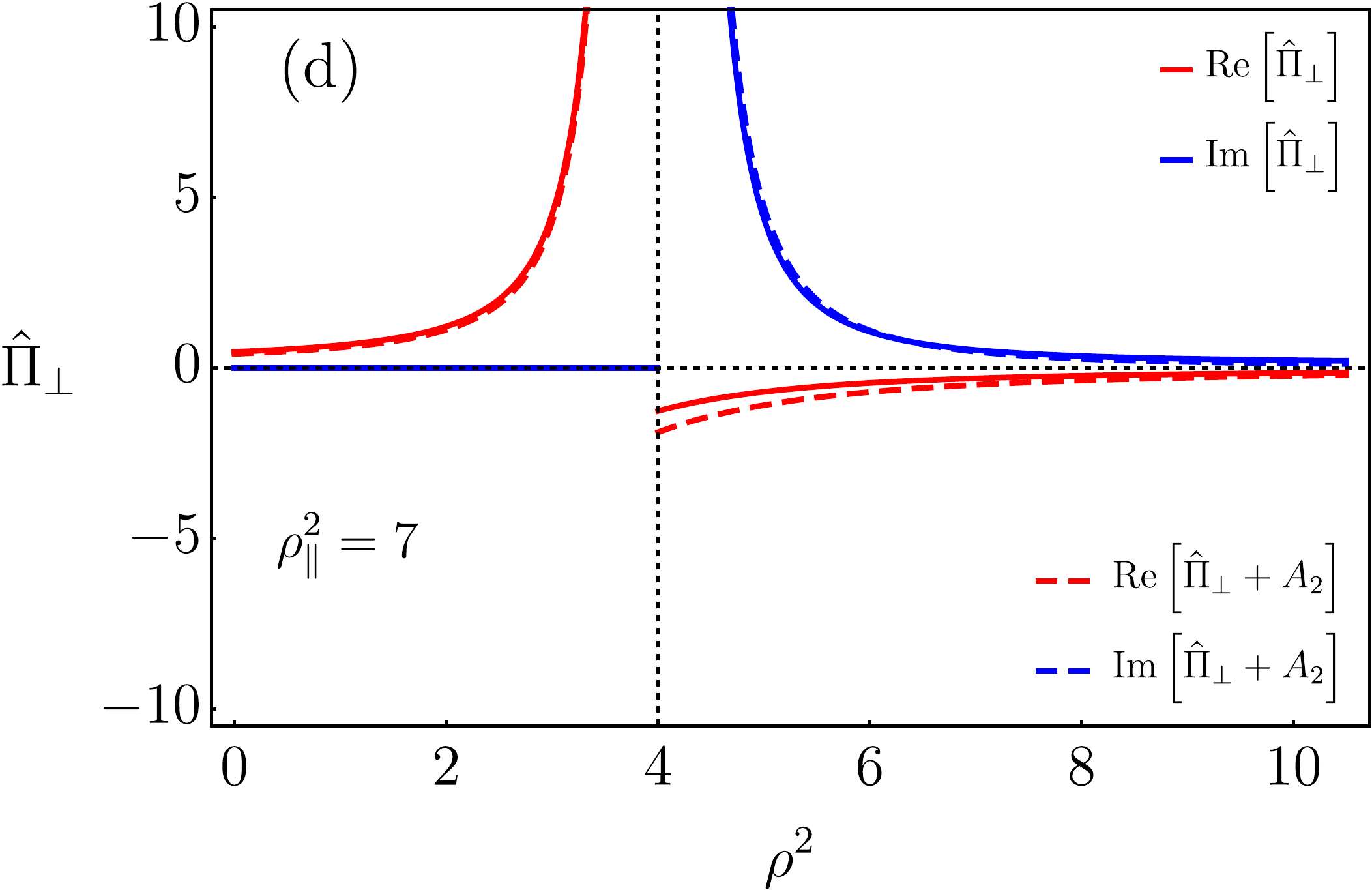}\\
    \vspace{0.5cm}
    \includegraphics[scale=0.4]{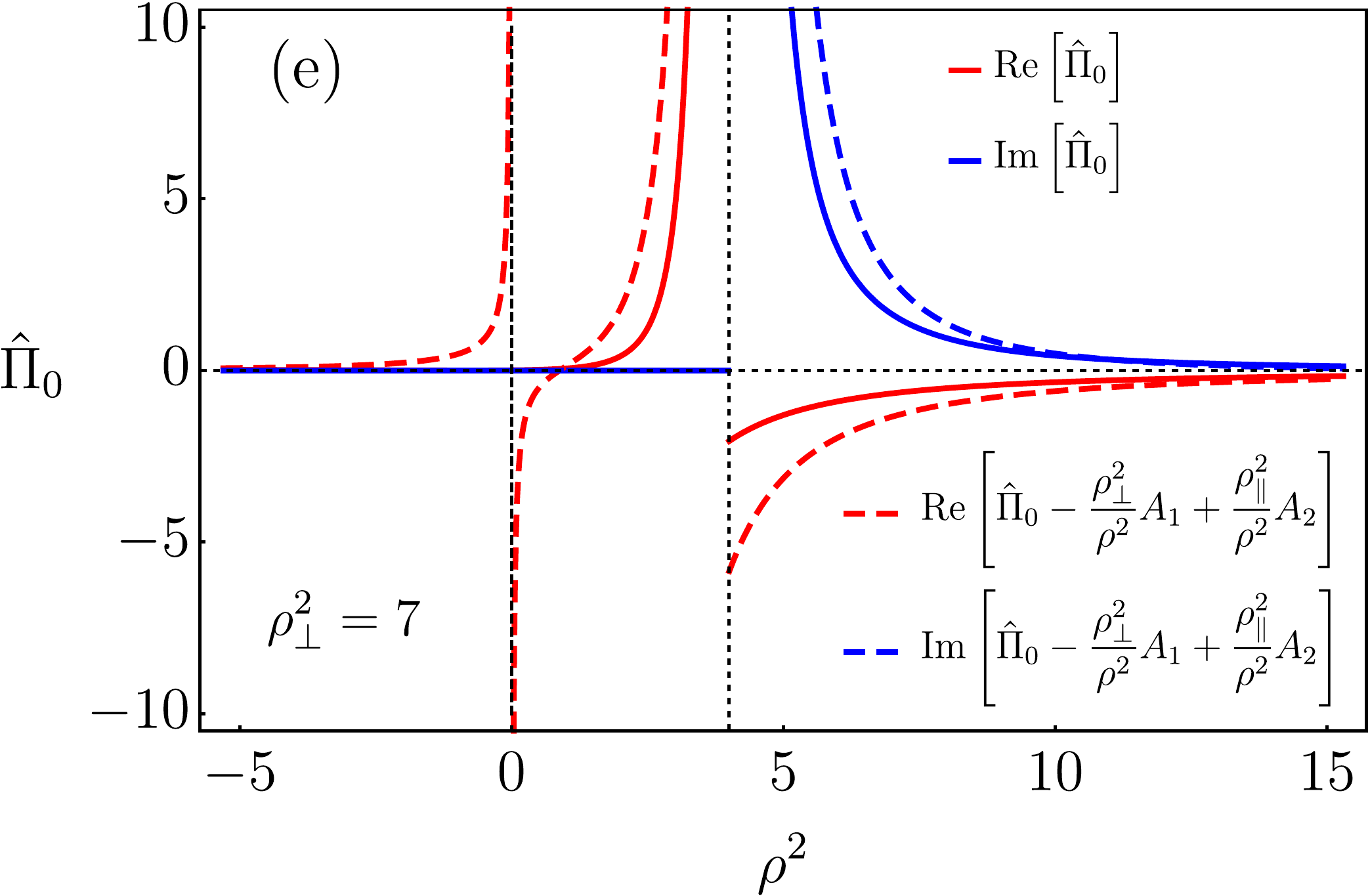}\hspace{0.6cm}\includegraphics[scale=0.4]{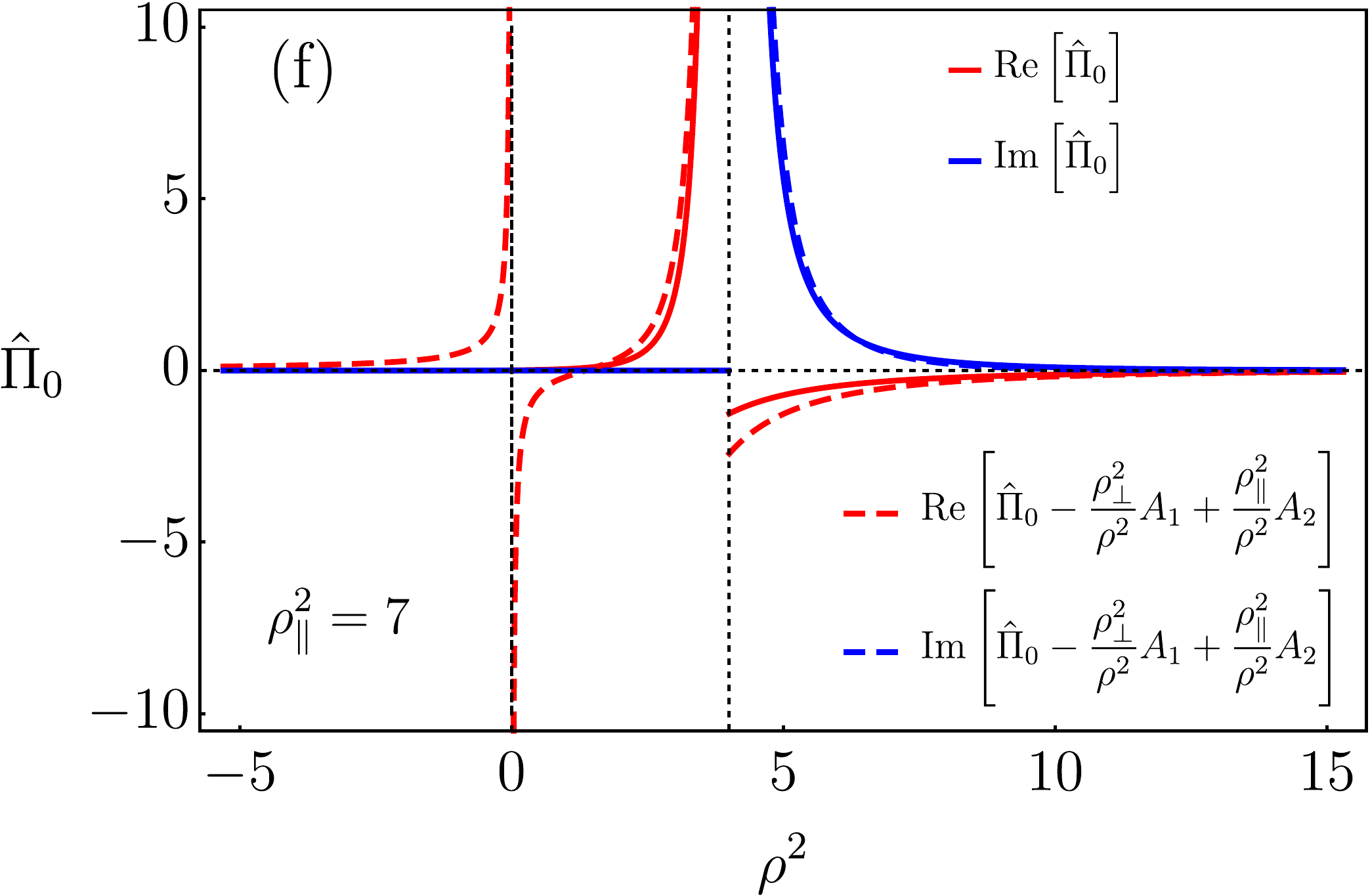}
    \caption{(Color online) Real and imaginary parts of the coefficients $\hat{\Pi}_\parallel,\,\hat{\Pi}_\perp$ and $\hat{\Pi}_0$ from Eqs.~(\ref{Piparaweakfield})-(\ref{Piceroweakfield}) as functions of $\rho^2$ for fixed values of $\rhop^2$ and $\rhop^2$. For comparison, these coefficients are also plotted including he spurious contributions from $A_1$ and $A_2$, given by Eqs.~(\ref{PiparaconA1})-(\ref{PiceroconA1A2}). Notice that for the chosen kinematical range for $\rho^2$, the threshold appears at $\rho^2=4$ or equivantly at $p^2=4m_f^2$, whereas the spurious terms contain unphysical thresholds at $\rho^2=0$.}
    \label{Fig:Coefweakfield}
\end{figure*}

Subtracting this contribution, we are left with the $\B$-dependent part. The integrations over $s$ and $y$ can be performed analytically, so that
\bea
&\!\!\!i&\!\!\!\Pi^{\mu\nu}_{\text{Weak}\;\B}
=-\frac{ig^2m_f^2\B^2}{6\pi^2}
\nonumber\\
&\times&\left[\hat{\Pi}_\parallel(\rho^2)\Pp+\hat{\Pi}_\perp(\rho^2)\Pt+\hat{\Pi}_0(\rho^2)\Pcero\right],
\eea
where
\begin{widetext}
\bea
\hat{\Pi}_\parallel&=&\frac{1}{4-\rho^2}\Bigg[\Bigg(12\frac{10+(\rho^2-6)\rho^2}{(4-\rho^2)^{3/2}(\rho^2)^{5/2}}\rhot^2+2\frac{\rho^2(\rho^2+2)-12}{\sqrt{4-\rho^2}(\rho^2)^{5/2}}\rhot^2+2\frac{(\rho^2-2)\sqrt{(4-\rho^2)\rho^2}}{(4-\rho^2)\rho^2}\Bigg)\arctan\left(\frac{\sqrt{\rho^2}}{\sqrt{4-\rho^2}}\right)\nn\\
&-&\frac{(\rho^2-10)(\rho^2-3)}{4-\rho^2}\frac{\rhot^2}{\rho^4}+\frac{6\rhot^2}{\rho^2}+1\Bigg],
\label{Piparaweakfield}
\eea

\bea
\hat{\Pi}_\perp&=&\frac{1}{4-\rho^2}\Bigg[\Bigg(12\frac{10+(\rho^2-6)\rho^2}{(4-\rho^2)^{3/2}(\rho^2)^{5/2}}\rhot^2+3\frac{(\rho^2-2)\sqrt{(4-\rho^2)\rho^2}}{(4-\rho^2)\rho^4}\rhot^2-\frac{\rho^2(\rho^2+2)-12}{\sqrt{4-\rho^2}(\rho^2)^{5/2}}\rhop^2\nn\\
&+&\frac{12+(\rho^2-6)\rho^2}{\sqrt{4-\rho^2}(\rho^2)^{3/2}}\Bigg)\arctan\left(\frac{\sqrt{\rho^2}}{\sqrt{4-\rho^2}}\right)\nn\\
&-&\frac{(\rho^2-10)(\rho^2-3)}{4-\rho^2}\frac{\rhot^2}{\rho^4}+\frac{3\rhot^2}{2\rho^2}-\frac{3\rhop^2}{\rho^4}+\frac{(\rho^2-3)\sqrt{(4-\rho^2)\rho^2}}{\sqrt{4-\rho^2}(\rho^2)^{3/2}}\Bigg],
\label{Piperpweakfield}
\eea

\bea
\hat{\Pi}_0&=&\frac{1}{4-\rho^2}\Bigg[\Bigg(12\frac{10+(\rho^2-6)\rho^2}{(4-\rho^2)^{3/2}(\rho^2)^{5/2}}\rhot^2+2\frac{12+(\rho^2-6)\rho^2}{\sqrt{4-\rho^2}(\rho^2)^{3/2}}-2\frac{(\rho^2-2)\sqrt{(4-\rho^2)\rho^2}}{(4-\rho^2)\rho^2}\Bigg)\arctan\left(\frac{\sqrt{\rho^2}}{\sqrt{4-\rho^2}}\right)\nn\\
&-&\frac{(\rho^2-10)(\rho^2-3)}{4-\rho^2}\frac{\rhot^2}{\rho^4}+2\frac{(\rho^2-3)\sqrt{(4-\rho^2)\rho^2}}{\sqrt{4-\rho^2}(\rho^2)^{3/2}}-1\Bigg]
\label{Piceroweakfield}
\eea

\end{widetext}
The coefficients $\hat{\Pi}_\parallel$, $\hat{\Pi}_\perp$ and $\hat{\Pi}_0$ consist of real and imaginary parts.
The imaginary parts can be obtained from the corresponding real parts from the Kramers-Kronig relations. With the notation $\omega \equiv \rho_0$, we have
\begin{eqnarray}
{\mbox{Im}}\Pi^{\mu\nu}(\omega) = -\frac{1}{\pi}\mathcal{P}\int_{-\infty}^{+\infty}\frac{{\mbox{Re}} \Pi^{\mu\nu}(\omega')}{\omega' - \omega}d\omega' ,
\end{eqnarray}
where $\mathcal{P}$ is the Principal Value. Examples of these coefficients as functions of $\rho_\parallel^2$, for various values of $\rho_\perp^2$ are shown in Fig.~\ref{Fig:Coefweakfield}. For comparison, these coefficients are also plotted including he spurious contributions from $A_1$ and $A_2$ given by Eqs.~(\ref{PiparaconA1})-(\ref{PiceroconA1A2}). Notice the appearance of unphysical thresholds at $\rho^2=0$ as well as large deviations from the correct functional behavior of the coefficients of the tensor structures.

\section{Results, discussion and conclusions}\label{V}

The results of this work can be used to study birefringence of the gluon polarization in a magnetized medium. Recall that
birefringence is the optical property exhibited by a material whose refractive index depends on the polarization and propagation direction of light. In solid-state, crystals with non-cubic lattice symmetry show birefringence, with calcite being a typical and historical example. The simplest type of birefringence corresponds to the so-called uniaxial type, where a single direction governs the optical anysotropy while all the other directions orthogonal to it are optically equivalent. Thus, rotations of the crystal with respect to this axis leave the optical response invariant. On the other hand, a material that is otherwise optically isotropic, can manifest birefringence under the presence of external agents, such as strain and, more importantly, an external magnetic field. This last case is often called Faraday effect \cite{Jain}. An analogous situation is studied in the context of high-energy physics, particularly in QED under the presence of static magnetic or electric fields, where the index of refraction depends on the photon polarization state. 

Despite the absence of an underlying discrete symmetry as in crystalline materials, the presence of these static fields is often sufficient to induce optical birefringence under certain conditions, which in this context is called vacuum birefringence. 
This effect has been extensively studied theoretically~\cite{Hattori,Dittrich_Reuter}. Moreover, recent experimental evidence for this phenomenon has been provided from astronomical observations of neutron stars, where intense magnetic fields are present~\cite{Mignani}. 

In QED, the microscopic mechanism behind the effect are the vacuum fluctuations due to the spontaneous emergence of virtual electron-positron pairs that act as dipoles, in analogy with dielectric crystals. In the absence of external fields, Lorentz invariance ensures an isotropic optical response. However, when a static electric or magnetic field is present, Lorentz invariance is broken and an anisotropic optical response is triggered. In particular, when a magnetic field is responsible for the effect, the virtual fermion pair exists in general in a  combination of  Landau levels. 

In this work, we show that vacuum birefringence  arises also for gluons in QCD, where the virtual fermion-antifermion pairs correspond to quark-antiquark pairs that play the same role as electron-positron pairs in QED. Just as in QED, the QCD version of the phenomenon necessarily implies the existence of an infinite sequence of momentum thresholds, that correspond to the condition where the external gluon momentum is resonant with the magnetic mass of a pair occupying a given Landau level, which are successively occupied by the pair of virtual quarks participating in the process. 

We have presented a method to compute the one-loop magnetic correction to the gluon polarization tensor starting from the Landau-level representation of the quark propagator in the presence of an external magnetic field. With suitable transformations, we have shown that this representation can be converted into the expression for the one-loop polarization tensor equivalent to the one obtained  starting from  Schwinger's proper time representation of the quark propagators. 
 We have shown that the general expression contains the vacuum contribution that can be isolated from the zero-field limit for finite gluon momentum. This can be achieved only when the whole sum over levels is performed. Therefore, calculations that resort to partial sums over Landau levels run the risk to mask the vacuum contributions and distort the result. An important observation is that, the general tensor structure for the gluon polarization contains two spurious terms that do not satisfy the transversality properties. We have shown that, in analogy with the case in vacuum, these terms have vanishing coefficients and thus do not contribute to the polarization tensor, as expected. Nevertheless, strictly speaking, this result requires that the $i\epsilon$ term in the quark propagators is not taken to zero, for otherwise the integrals representing the coefficients of the spurious terms are not oscillatory and the areas above and below the $x_i$ axis cannot cancel. However, as it is customary, this term is only kept implicit in the calculation and is only brought back, for instance, when computing the real and imaginary parts of the final result. Thus, if the coefficients of the spurious terms are not shown to vanish and care is not taken when computing the coefficients upon projection onto the chosen basis, these can give contributions that are not correct, as we have shown. In order to check the validity of the expressions thus found, we have shown that the strong field limit obtained from our approach reproduces a well established result. The results of this work can be used to study the conditions for gluons to equilibrate with a magnetized medium, for example during the early stages of a relativistic heavy-ion collision. This is work in progress and it will be reported elsewhere.

\section*{Acknowledgements}
The authors are thankful to L. A. Hern\'andez and R. Zamora for useful conversations. Support for
this work was received in part by UNAM-DGPA-PAPIIT grant number IG100219 and by Consejo Nacional de Ciencia y Tecnolog\'ia grant number 256494. E. M. acknowledges support from FONDECYT (Chile) under grant No. 1190361. M. Loewe acknowledges support from FONDECYT (Chile) under grants No. 1170107, 1190192 and from Conicyt/PIA/Basal (Chile) grant number FB0821.

\begin{widetext}

\newpage
\appendix

\section{Derivation of Eqs.~(\ref{fs})}\label{ApA}
Let us begin from the general expression of the gluon polarization tensor of Eq.~(\ref{Pidef}):
\bea
i\Pi^{\mu\nu}_{(ab)}&=&-\frac{1}{2}\int\frac{d^4k}{\dpi^4}\text{Tr}\left\{igt_b\gamma^\nu iS^{(n)}(k)igt_a\gamma^\mu iS^{(m)}(q)\right\}+ {\mbox{C.C.}}.
\eea
The trace in the above expression involves two  fermion propagator factors, each given by Eqs.~(\ref{fermionpropdef})-(\ref{Dn}). This product produces nine terms, that are explicitly given by
\bea
t_1^{\mu\nu}&=&-g^2\sum_{n,m=0}^{\infty}\int\frac{d^4k}{\dpi^4}\exp\left[-\frac{\kt^2+(k-p)_{\perp}^2}{\eB}\right]\frac{(-1)^{n+m}L_n^0\left(\frac{2\kt^2}{\eB}\right)L_{m}^0\left[\frac{2(k-p)_{\perp}^2}{\eB}\right]}{\left[k_\p^2-m_f^2-2n\left |  q_fB\right |\right]\left[(k-p)_{\parallel}^2-m_f^2-2m\left |  q_fB\right |\right]}\nn\\
&\times&\Tr\left\{\gn(\ks_\p+m_f)\Op^{-}\gm(\ks_\p-\ps_\p+m_f)\Op^{-}\right\}+\text{C.C.}
\label{t1}
\eea

\bea
t_2^{\mu\nu}&=&g^2\sum_{n=0,m=1}^{\infty}\int\frac{d^4k}{\dpi^4}\exp\left[-\frac{\kt^2+(k-p)_{\perp}^2}{\eB}\right]\frac{(-1)^{n+m}L_n^0\left(\frac{2\kt^2}{\eB}\right)L_{m-1}^0\left[\frac{2(k-p)_{\perp}^2}{\eB}\right]}{\left[k_\p^2-m_f^2-2n\left |  q_fB\right |\right]\left[(k-p)_{\parallel}^2-m_f^2-2m\left |  q_fB\right |\right]}\nn\\
&\times&\Tr\left\{\gn(\ks_\p+m_f)\Op^{-}\gm(\ks_\p-\ps_\p+m_f)\Op^{+}\right\}+\text{C.C.}
\label{t2}
\eea

\bea
t_3^{\mu\nu}&=&-2g^2\sum_{n=0,m=1}^{\infty}\int\frac{d^4k}{\dpi^4}\exp\left[-\frac{\kt^2+(k-p)_{\perp}^2}{\eB}\right]\frac{(-1)^{n+m}L_n^0\left(\frac{2\kt^2}{\eB}\right)L_{m-1}^1\left[\frac{2(k-p)_{\perp}^2}{\eB}\right]}{\left[k_\p^2-m_f^2-2n\left |  q_fB\right |\right]\left[(k-p)_{\parallel}^2-m_f^2-2m\left |  q_fB\right |\right]}\nn\\
&\times&\Tr\left\{\gn(\ks_\p+m_f)\Op^{-}\gm(\ks_\perp-\ps_\perp)\right\}+\text{C.C.}
\label{t3}
\eea

\bea
t_4^{\mu\nu}&=&g^2\sum_{n=1,m=0}^{\infty}\int\frac{d^4k}{\dpi^4}\exp\left[-\frac{\kt^2+(k-p)_{\perp}^2}{\eB}\right]\frac{(-1)^{n+m}L_{n-1}^0\left(\frac{2\kt^2}{\eB}\right)L_{m}^0\left[\frac{2(k-p)_{\perp}^2}{\eB}\right]}{\left[k_\p^2-m_f^2-2n\left |  q_fB\right |\right]\left[(k-p)_{\parallel}^2-m_f^2-2m\left |  q_fB\right |\right]}\nn\\
&\times&\Tr\left\{\gn(\ks_\p+m_f)\Op^{+}\gm(\ks_\p-\ps_\p+m_f)\Op^{-}\right\}+\text{C.C.}
\label{t4}
\eea

\bea
t_5^{\mu\nu}&=&-g^2\sum_{n=1,m=1}^{\infty}\int\frac{d^4k}{\dpi^4}\exp\left[-\frac{\kt^2+(k-p)_{\perp}^2}{\eB}\right]\frac{(-1)^{n+m}L_{n-1}^0\left(\frac{2\kt^2}{\eB}\right)L_{m-1}^0\left[\frac{2(k-p)_{\perp}^2}{\eB}\right]}{\left[k_\p^2-m_f^2-2n\left |  q_fB\right |\right]\left[(k-p)_{\parallel}^2-m_f^2-2m\left |  q_fB\right |\right]}\nn\\
&\times&\Tr\left\{\gn(\ks_\p+m_f)\Op^{+}\gm(\ks_\p-\ps_\p+m_f)\Op^{+}\right\}+\text{C.C.}
\label{t5}
\eea

\bea
t_6^{\mu\nu}&=&2g^2\sum_{n=1,m=1}^{\infty}\int\frac{d^4k}{\dpi^4}\exp\left[-\frac{\kt^2+(k-p)_{\perp}^2}{\eB}\right]\frac{(-1)^{n+m}L_{n-1}^0\left(\frac{2\kt^2}{\eB}\right)L_{m-1}^1\left[\frac{2(k-p)_{\perp}^2}{\eB}\right]}{\left[k_\p^2-m_f^2-2n\left |  q_fB\right |\right]\left[(k-p)_{\parallel}^2-m_f^2-2m\left |  q_fB\right |\right]}\nn\\
&\times&\Tr\left\{\gn(\ks_\p+m_f)\Op^{+}\gm(\ks_\perp-\ps_\perp)\right\}+\text{C.C.}
\label{t6}
\eea

\bea
t_7^{\mu\nu}&=&-2g^2\sum_{n=1,m=0}^{\infty}\int\frac{d^4k}{\dpi^4}\exp\left[-\frac{\kt^2+(k-p)_{\perp}^2}{\eB}\right]\frac{(-1)^{n+m}L_{n-1}^1\left(\frac{2\kt^2}{\eB}\right)L_{m}^0\left[\frac{2(k-p)_{\perp}^2}{\eB}\right]}{\left[k_\p^2-m_f^2-2n\left |  q_fB\right |\right]\left[(k-p)_{\parallel}^2-m_f^2-2m\left |  q_fB\right |\right]}\nn\\
&\times&\Tr\left\{\gn\ks_\perp\gm(\ks_\p-\ps_\p+m_f)\Op^{-}\right\}+\text{C.C.}
\label{t7}
\eea

\bea
t_8^{\mu\nu}&=&2g^2\sum_{n=1,m=1}^{\infty}\int\frac{d^4k}{\dpi^4}\exp\left[-\frac{\kt^2+(k-p)_{\perp}^2}{\eB}\right]\frac{(-1)^{n+m}L_{n-1}^1\left(\frac{2\kt^2}{\eB}\right)L_{m-1}^0\left[\frac{2(k-p)_{\perp}^2}{\eB}\right]}{\left[k_\p^2-m_f^2-2n\left |  q_fB\right |\right]\left[(k-p)_{\parallel}^2-m_f^2-2m\left |  q_fB\right |\right]}\nn\\
&\times&\Tr\left\{\gn\ks_\perp\gm(\ks_\p-\ps_\p+m_f)\Op^{+}\right\}+\text{C.C.}
\label{t8}
\eea

\bea
t_9^{\mu\nu}&=&4g^2\sum_{n=1,m=1}^{\infty}\int\frac{d^4k}{\dpi^4}\exp\left[-\frac{\kt^2+(k-p)_{\perp}^2}{\eB}\right]\frac{(-1)^{n+m}L_{n-1}^1\left(\frac{2\kt^2}{\eB}\right)L_{m-1}^1\left[\frac{2(k-p)_{\perp}^2}{\eB}\right]}{\left[k_\p^2-m_f^2-2n\left |  q_fB\right |\right]\left[(k-p)_{\parallel}^2-m_f^2-2m\left |  q_fB\right |\right]}\nn\\
&\times&\Tr\left\{\gn\ks_\perp\gm(\ks_\perp-\ps_\perp)\right\}+\text{C.C.}
\label{t9}
\eea
In order to perform the sum over Landau levels, we write the denominators introducing Schwinger parameters such that
\bea
\frac{1}{y}=\int_0^\infty e^{-y x} dx.
\eea
We start with the expression given by Eq.~(\ref{t1})
\bea
t_1^{\mu\nu}&=&-g^2\int\frac{d^4k}{\dpi^4}\exp\left[-\frac{\kt^2+(k-p)_{\perp}^2}{\eB}\right]\Tr\left\{\gn(\ks_\p+m_f)\Op^{-}\gm(\ks_\p-\ps_\p+m_f)\Op^{-}\right\}\nn\\
&\times&\sum_{n,m=0}^{\infty}\frac{(-1)^{n+m}L_n^0\left(\frac{2\kt^2}{\eB}\right)L_{m}^0\left[\frac{2(k-p)_{\perp}^2}{\eB}\right]}{\left[k_\p^2-m_f^2-2n\left |  q_fB\right |\right]\left[(k-p)_{\parallel}^2-m_f^2-2m\left |  q_fB\right |\right]}\nn\\
&=&-g^2\int d^2x\int\frac{d^4k}{\dpi^4}\exp\left[-\frac{\kt^2+(k-p)_{\perp}^2}{\eB}\right]\Tr\left\{\gn(\ks_\p+m_f)\Op^{-}\gm(\ks_\p-\ps_\p+m_f)\Op^{-}\right\}\nn\\
&\times&e^{\alpha(\kp)x_1+\beta(\kp)x_2}\sum_{n,m=0}^{\infty}r_1^nL_n^0(s_1)r_2^mL_m^0(s_2)+\text{C.C.}
\eea
where
\begin{subequations}
\bea
\alpha(\kp)&=&\kp^2-m_f^2,
\label{alphadef}
\eea

\bea
\beta(\kp)&=&(\kp-p_\p)^2-m_f^2,
\label{betadef}
\eea

\bea
r_i&=&-e^{-2\eB x_i},\; i=1,2,
\label{ri}
\eea

\bea
s_1=\frac{2\kt^2}{\eB},
\label{s1}
\eea

and
\bea
s_2=\frac{2(k-p)_{\perp}^2}{\eB}.
\label{s2}
\eea
\end{subequations}
By using the generating function of the Laguerre Polynomials, given by
\bea
\sum_{n=0}^\infty r^n L_n^{b}(s)=\frac{1}{\left(1-r\right)^{b+1}}\exp\left(-\frac{r}{1-r}s\right),
\label{GeneratingFunction}
\eea
we find
\bea
t_1^{\mu\nu}&=&-g^2\int \frac{d^2x}{\left(1+e^{-2\eB x_1}\right)\left(1+e^{-2\eB x_2}\right)}\int\frac{d^4k}{\dpi^4}\exp\left[-\frac{\kt^2+(k-p)_{\perp}^2}{\eB}\right]e^{\alpha(\kp)x_1+\beta(\kp)x_2}\nn\\
&\times&\exp\left[ n(x_1)\frac{2\kt^2}{\eB}\right]\exp\left[ n(x_2)\frac{2(k-p)_{\perp}^2}{\eB}\right]\Tr\left\{\gn(\ks_\p+m_f)\Op^{-}\gm(\ks_\p-\ps_\p+m_f)\Op^{-}\right\}+\text{C.C.}
\eea
where we have defined
\bea
n(x)\equiv\frac{1}{e^{2\eB x}+1}.
\label{ndef}
\eea
 Now, for the trace computation, note that
 \begin{subequations}
\bea
\left[\gm_\p,\Op^{(\pm)}\right]=0,
\label{Ogammaconmmutator}
\eea

\bea
\Op^{(\pm)}\gm\Op^{(\pm)}=\Op^{(\pm)}\gm_\p,
\label{Ogammapara}
\eea
\end{subequations}
and therefore
\bea
&&4\Tr\left\{\gn(\ks_\p+m_f)\gm_\p(\ks_\p-\ps_\p+m_f)\Op^{-}\right\}+\text{C.C.}\nn\\
&=&4\Tr\left\{\gn(\ks_\p+m_f)\gm_\p(\ks_\p-\ps_\p+m_f)\right\}\nn\\
&=&16\left[\left(\kp\cdot\pp+m_f^2-\kp^2\right)\gmn_\p+2\km_\p\kn_\p-\km_\p p^\nu_\p-\kn_\p p^\mu_\p\right].
\eea
Putting all together
\bea
t_1^{\mu\nu}&=&-4g^2\int \frac{\mathcal{I}_1(x_1,x_2)\mathcal{J}_1^{\mu\nu}(x_1,x_2)}{\left(1+e^{-2\eB x_1}\right)\left(1+e^{-2\eB x_2}\right)}d^2x,\nn\\
\eea
with
\bea
\mathcal{I}_1&=&\int\frac{d^2k_\perp}{\dpi^2}\exp\left[-\frac{\kt^2+(k-p)_{\perp}^2}{\eB}\right]\exp\left[ n(x_1)\frac{2\kt^2}{\eB}\right]\exp\left[ n(x_2)\frac{2(k-p)_{\perp}^2}{\eB}\right]
\label{I1}
\eea
and
\bea
&&\mathcal{J}_1^{\mu\nu}=\int\frac{d^2\kp}{\dpi^2}e^{\alpha(\kp)x_1}e^{\beta(\kp)x_2}\left[\left(\kp\cdot\pp+m_f^2-\kp^2\right)\gmn_\p+2\km_\p\kn_\p-\km_\p p^\nu_\p-\kn_\p p^\mu_\p\right].
\label{J1}
\eea

The transverse integral $\mathcal{I}_1$ is performed by making the shift
\bea
\kt=\qt+\frac{1-2n(x_2)}{2\left[1-n(x_1)-n(x_2)\right]}\pt,
\label{qtshift}
\eea
which turns the integral into a simple Gaussian form. It is straightforward to prove that
\bea
\mathcal{I}_1&=&\frac{\pi}{\dpi^2}\frac{\eB}{\tanh\left(\eB x_1\right)+\tanh\left(\eB x_2\right)}\exp\left[-\frac{\tanh(\eB x_1)\tanh(\eB x_2)}{\tanh(\eB x_1)+\tanh(\eB x_2)}\frac{\pt^2}{\eB}\right].
\label{I1result}
\eea

For the parallel integral $\mathcal{J}_1^{\mu\nu}$, the appropriate shift is
\bea
l=\kp-\frac{x_2}{x_1+x_2}\pp
\label{ldef},
\eea
and by performing a rotation to Euclidean space, the integral becomes of a Gaussian form in the variable $l^2_E=l_4^2+l_3^2$, and thus
\bea
&&\mathcal{J}_1^{\mu\nu}=\frac{i\pi}{\dpi^2}\exp\left[\frac{x_1x_2}{x_1+x_2}\pp^2-m_f^2(x_1+x_2)\right]\left[\left(\frac{x_1x_2}{(x_1+x_2)^3}\pp^2+\frac{m_f^2}{x_1+x_2}\gmnp\right)-\frac{2x_1x_2}{(x_1+x_2)^3}\pp^\mu\pp^\nu\right].
\eea
Collecting terms
\bea
&&t_1^{\mu\nu}=-\frac{i\eB}{16\pi^2}g^2\int d^2x\frac{e^{\eB(x_1+x_2)}}{\sinh\left[\eB(x_1+x_2)\right]}\exp\left[\frac{x_1x_2}{x_1+x_2}\pp^2-m_f^2(x_1+x_2)\right]\nn\\
&\times&\exp\left[-\frac{\tanh(\eB x_1)\tanh(\eB x_2)}{\tanh(\eB x_1)+\tanh(\eB x_2)}\frac{\pt^2}{\eB}\right]\left[\left(\frac{x_1x_2}{(x_1+x_2)^3}\pp^2+\frac{m_f^2}{x_1+x_2}\right)\gmnp-\frac{2x_1x_2}{(x_1+x_2)^3}\pp^\mu\pp^\nu\right].
\eea

Note that the term $t_5^{\mu\nu}$ of Eq. (\ref{t5}) has the same tensor structure as $t_1^{\mu\nu}$. By means of the variable shifts $m'=m-1$ and $n'=n-1$, which produce a factor $e^{-2\eB (x_1+x_2)}$, this gives rise at the same set of transverse and parallel integrals than for the case of $t_1^{\mu\nu}$. Therefore, we can write
\bea
&&t_5^{\mu\nu}=-\frac{i\eB}{16\pi^2}g^2\int d^2x\frac{e^{-\eB(x_1+x_2)}}{\sinh\left[\eB(x_1+x_2)\right]}\exp\left[\frac{x_1x_2}{x_1+x_2}\pp^2-m_f^2(x_1+x_2)\right]\nn\\
&\times&\exp\left[-\frac{\tanh(\eB x_1)\tanh(\eB x_2)}{\tanh(\eB x_1)+\tanh(\eB x_2)}\frac{\pt^2}{\eB}\right]\left[\left(\frac{x_1x_2}{(x_1+x_2)^3}\pp^2+\frac{m_f^2}{x_1+x_2}\right)\gmnp-\frac{2x_1x_2}{(x_1+x_2)^3}\pp^\mu\pp^\nu\right].
\eea
Adding up these two terms, we get
\bea
&&t_1^{\mu\nu}+t_5^{\mu\nu}=-\frac{i\eB}{8\pi^2}g^2\int d^2x\coth\left[\eB(x_1+x_2)\right]\exp\left[\frac{x_1x_2}{x_1+x_2}\pp^2-m_f^2(x_1+x_2)\right]\nn\\
&\times&\exp\left[-\frac{\tanh(\eB x_1)\tanh(\eB x_2)}{\tanh(\eB x_1)+\tanh(\eB x_2)}\frac{\pt^2}{\eB}\right]\left[\left(\frac{x_1x_2}{(x_1+x_2)^3}\pp^2+\frac{m_f^2}{x_1+x_2}\right)\gmnp-\frac{2x_1x_2}{(x_1+x_2)^3}\pp^\mu\pp^\nu\right].\nn\\
&\equiv&\factorglobal f_0(x_1,x_2)f_1^{\mu\nu}(x_1,x_2).
\eea
For the term $t_2^{\mu\nu}$ of Eq.~(\ref{t2}) the trace involved is computed by using Eq.~(\ref{Ogammaconmmutator}) and the relation
\bea
\Op^{(\pm)}\gm\Op^{(\mp)}=\Op^{(\pm)}\gm_\perp,
\label{Ogammaperp}
\eea
so that
\bea
\Tr\left\{\gn_\perp(\ks_\p+m_f)\gm(\ks_\p-\ps_\p+m_f)\right\}+\text{C.C.}=4\left(\kp\cdot\pp-\kp^2+m_f^2\right)\gmn_\perp.
\eea
This results imply that after introducing the Schwinger parametrization, the integration over the transverse momentum gives the same results as those in Eq.~(\ref{I1result}). Moreover, in order to apply  Eq.~(\ref{GeneratingFunction}) it is necessary to perform the shift $m'=m-1$. That shift implies extracting a factor $-e^{-2\eB x_2}$ from the sum , thus
\bea
t_2^{\mu\nu}&=&-\frac{4\pi\eB}{\dpi^4}g^2\int \,d^2x \frac{e^{-2\eB x_2}}{\left(1+e^{-2\eB x_1}\right)\left(1+e^{-2\eB x_2}\right)}\frac{\eB}{\tanh\left(\eB x_1\right)+\tanh\left(\eB x_2\right)}\nn\\
&\times&\exp\left[-\frac{\tanh(\eB x_1)\tanh(\eB x_2)}{\tanh(\eB x_1)+\tanh(\eB x_2)}\frac{\pt^2}{\eB}\right]\int d^2\kp\left(\kp\cdot\pp-\kp^2+m_f^2\right)e^{\alpha(\kp)x_1}e^{\beta(\kp)x_2}g^{\mu\nu}_\perp.
\eea
The parallel integration is carried out with the help of the momentum shift of Eq.~(\ref{ldef}) which in Euclidean space gives
\bea
t_2^{\mu\nu}&=&-\frac{4i\pi^2\eB}{\dpi^2}g^2\int \,d^2x \frac{e^{-2\eB x_2}}{\left(1+e^{-2\eB x_1}\right)\left(1+e^{-2\eB x_2}\right)}\frac{\eB}{\tanh\left(\eB x_1\right)+\tanh\left(\eB x_2\right)}\nn\\
&\times&\exp\left[\frac{x_1x_2}{x_1+x_2}\pp^2-m_f^2(x_1+x_2)\right]\exp\left[-\frac{\tanh(\eB x_1)\tanh(\eB x_2)}{\tanh(\eB x_1)+\tanh(\eB x_2)}\frac{\pt^2}{\eB}\right]\nn\\
&\times&\left[\frac{x_1x_2}{(x_1+x_2)^3}\pp^2+\frac{m_f^2}{x_1+x_2}+\frac{1}{(x_1+x_2)^2}\right]\gmn_\perp.
\eea
From the fact that the term $t_4^{\mu\nu}$ has the same tensor structure of $t_1^{\mu\nu}$, it is easy to show that both expressions are related to each other after the exchange $x_1\leftrightarrow x_2$, so that
\bea
t_4^{\mu\nu}&=&-\frac{4i\pi^2\eB}{\dpi^2}g^2\int \,d^2x \frac{e^{-2\eB x_1}}{\left(1+e^{-2\eB x_1}\right)\left(1+e^{-2\eB x_2}\right)}\exp\left[\frac{x_1x_2}{x_1+x_2}\pp^2-m_f^2(x_1+x_2)\right]\nn\\
&\times&\exp\left[-\frac{\tanh(\eB x_1)\tanh(\eB x_2)}{\tanh(\eB x_1)+\tanh(\eB x_2)}\frac{\pt^2}{\eB}\right]\left[\frac{x_1x_2}{(x_1+x_2)^3}\pp^2+\frac{m_f^2}{x_1+x_2}+\frac{1}{(x_1+x_2)^2}\right]\gmn_\perp,
\eea
 and therefore, after manipulating the exponential, we get
\bea
&&t_2^{\mu\nu}+t_4^{\mu\nu}=-\frac{i\eB}{8\pi^2}g^2\int d^2x\frac{\cosh\left[\eB(x_2-x_1)\right]}{\sinh\left[\eB(x_1+x_2)\right]}\exp\left[\frac{x_1x_2}{x_1+x_2}\pp^2-m_f^2(x_1+x_2)\right]\nn\\
&\times&\exp\left[-\frac{\tanh(\eB x_1)\tanh(\eB x_2)}{\tanh(\eB x_1)+\tanh(\eB x_2)}\frac{\pt^2}{\eB}\right]\left[\frac{x_1x_2}{(x_1+x_2)^3}\pp^2+\frac{m_f^2}{x_1+x_2}+\frac{1}{(x_1+x_2)^2}\right]\gmn_\perp\nn\\
&\equiv&\factorglobal f_0(x_1,x_2)f_2^{\mu\nu}(x_1,x_2).
\eea
For the term $t_3^{\mu\nu}$, the trace is computed with the help of Eqs.~(\ref{Ogammaconmmutator}),(\ref{Ogammapara}) and (\ref{Ogammaperp})
\bea
&&\Tr\left\{\gn(\ks_\p+m_f)\Op^{-}\gm(\ks_\perp-\ps_\perp)\right\}+\text{C.C.}=4\left[\kp^\mu\left(\kt^\nu-\pt^\nu\right)+\kp^\nu\left(\kt^\mu-\pt^\mu\right)\right].
\eea
After introducing Schwinger's parametrization and using the generating function for the Laguerre polynomials (with the shift $m'=m-1$), we obtain
\bea
t_3^{\mu\nu}&=&-\frac{8}{\dpi^4}g^2\int d^2x\int d^4k\frac{e^{-2\eB x_2}e^{\alpha(\kp)x_1}e^{\beta(\kp)x_2}}{\left(1+2e^{-2\eB x_1}\right)\left(1+2e^{-2\eB x_1}\right)^2}\exp\left[-\frac{\kt^2+(k-p)_{\perp}^2}{\eB}\right]\nn\\
&\times&\exp\left[ n(x_1)\frac{2\kt^2}{\eB}\right]\exp\left[ n(x_2)\frac{2(k-p)_{\perp}^2}{\eB}\right]\left[\kp^\mu\left(\kt^\nu-\pt^\nu\right)+\kp^\nu\left(\kt^\mu-\pt^\mu\right)\right].
\eea
The change of variable in Eq.~(\ref{qtshift}) leads to the result
\bea
t_3^{\mu\nu}&=&-\frac{8}{\dpi^4}g^2\int d^2x\int d^4k\frac{e^{-2\eB x_2}\left[\mathcal{I}_2^{\mu\nu}(x_1,x_2)+\mathcal{I}_2^{\nu\mu}(x_1,x_2)\right]}{\left(1+2e^{-2\eB x_1}\right)\left(1+2e^{-2\eB x_1}\right)^2}\exp\left[-\frac{\tanh(\eB x_1)\tanh(\eB x_2)}{\tanh(\eB x_1)+\tanh(\eB x_2)}\frac{\pt^2}{\eB}\right],\nn\\
\eea
where
\bea
\mathcal{I}_2^{\mu\nu}&=&\int d^2\kp e^{\alpha(\kp)x_1}e^{\beta(\kp)x_2}\int d^2q_\perp e^{-\eta\,\qt^2}\kp^\mu\left[\qt^\nu+(\sigma-1)\pt^\nu\right],
\eea
with
\begin{subequations}
\bea
\eta\equiv\frac{\tanh(\eB x_1)+\tanh(\eB x_2)}{\eB},
\eea
and
\bea
\sigma\equiv\frac{\tanh(\eB x_2)}{\tanh(\eB x_1)+\tanh(\eB x_2)}.
\eea
\label{etaandsigma}
\end{subequations}
The perpendicular integration has a simple Gaussian form for which the linear term in $\qt$ integrates to zero, yielding
\bea
\mathcal{I}_2^{\mu\nu}&=&\frac{\pi(\sigma-1)}{\eta}\pt^\nu\int d^2\kp e^{\alpha(\kp)x_1}e^{\beta(\kp)x_2}\kp^\mu.\nn\\
\eea
The shift of variable in Eq.~(\ref{ldef}) also implies a Gaussian integration (in Euclidean space), where the linear terms in $l$ vanish after integration. In this way
\bea
\mathcal{I}_2^{\mu\nu}&=&\frac{\pi^2(\sigma-1)}{\eta}\frac{x_2}{(x_1+x_2)^2}\pp^\mu\pt^\nu\exp\left[\frac{x_1x_2}{x_1+x_2}\pp^2-m_f^2(x_1+x_2)\right]\nn\\
&=&-\frac{i\pi^2\eB x_2}{(x_1+x_2)^2}\frac{\tanh(\eB x_1)\pp^\mu\pt^\nu}{\left[\tanh(\eB x_1)+\tanh(\eB x_2)\right]^2}.
\eea
Putting together these results
\bea
t_3^{\mu\nu}&=&-\frac{i\eB\pi^2}{\dpi^4}g^2\int d^2x\frac{x_2 e^{\eB x_1}\sinh(\eB x_1)}{(x_1+x_2)^2\sinh^2\left[\eB (x_1+x_2)\right]}\exp\left[\frac{x_1x_2}{x_1+x_2}\pp^2-m_f^2(x_1+x_2)\right]\nn\\
&\times&\exp\left[-\frac{\tanh(\eB x_1)\tanh(\eB x_2)}{\tanh(\eB x_1)+\tanh(\eB x_2)}\frac{\pt^2}{\eB}\right]\left(\pp^\mu\pt^\nu+\pp^\nu\pt^\mu\right).
\eea
The structure $t_6^{\mu\nu}$ is obtained from $t_3^{\mu\nu}$ after the shift $n'=n-1$ wichs means introducing a factor $-e^{-2\eB x_1}$, thus
\bea
t_6^{\mu\nu}&=&-\frac{i\eB\pi^2}{\dpi^4}g^2\int d^2x\frac{x_2 e^{-\eB x_1}\sinh(\eB x_1)}{(x_1+x_2)^2\sinh^2\left[\eB (x_1+x_2)\right]}\exp\left[\frac{x_1x_2}{x_1+x_2}\pp^2-m_f^2(x_1+x_2)\right]\nn\\
&\times&\exp\left[-\frac{\tanh(\eB x_1)\tanh(\eB x_2)}{\tanh(\eB x_1)+\tanh(\eB x_2)}\frac{\pt^2}{\eB}\right]\left(\pp^\mu\pt^\nu+\pp^\nu\pt^\mu\right),
\eea
and therefore
\bea
&&t_3^{\mu\nu}+t_6^{\mu\nu}=-\frac{i\eB}{8\pi^2}g^2\int d^2x\frac{x_2 \cosh(\eB x_1)\sinh(\eB x_1)}{(x_1+x_2)^2\sinh^2\left[\eB (x_1+x_2)\right]}\exp\left[\frac{x_1x_2}{x_1+x_2}\pp^2-m_f^2(x_1+x_2)\right]\nn\\
&\times&\exp\left[-\frac{\tanh(\eB x_1)\tanh(\eB x_2)}{\tanh(\eB x_1)+\tanh(\eB x_2)}\frac{\pt^2}{\eB}\right]\left(\pp^\mu\pt^\nu+\pp^\nu\pt^\mu\right).
\eea
Coming now to the terms $t_7^{\mu\nu}$ and $t_8^{\mu\nu}$, we notice that they share a common tensor form. Starting from $t_3^{\mu\nu}$, the expression for $t_7^{\mu\nu}$ is obtained by replacing $x_1\rightarrow x_2$ and $p\rightarrow-p$. Moreover, $t_8^{\mu\nu}$ is obtained from $t_7^{\mu\nu}$ by performing the shift $m'=m-1$ which amounts to introducing a factor $-e^{-2\eB x_2}$. Implementing these observations, we get
\bea
&&t_7^{\mu\nu}+t_8^{\mu\nu}=-\frac{i\eB}{8\pi^2}g^2\int d^2x\frac{x_1 \cosh(\eB x_2)\sinh(\eB x_2)}{(x_1+x_2)^2\sinh^2\left[\eB (x_1+x_2)\right]}\exp\left[\frac{x_1x_2}{x_1+x_2}\pp^2-m_f^2(x_1+x_2)\right]\nn\\
&\times&\exp\left[-\frac{\tanh(\eB x_1)\tanh(\eB x_2)}{\tanh(\eB x_1)+\tanh(\eB x_2)}\frac{\pt^2}{\eB}\right]\left(\pp^\mu\pt^\nu+\pp^\nu\pt^\mu\right),
\eea
then
\bea
&&t_3^{\mu\nu}+t_6^{\mu\nu}+t_7^{\mu\nu}+t_8^{\mu\nu}=-\frac{i\eB}{8\pi^2}g^2\int \frac{d^2x}{2(x_1+x_2)^2\sinh^2\left[\eB (x_1+x_2)\right]}\exp\left[\frac{x_1x_2}{x_1+x_2}\pp^2-m_f^2(x_1+x_2)\right]\nn\\
&\times&\exp\left[-\frac{\tanh(\eB x_1)\tanh(\eB x_2)}{\tanh(\eB x_1)+\tanh(\eB x_2)}\frac{\pt^2}{\eB}\right]\Big[x_1\sinh(2\eB x_2)+x_2\sinh(2\eB x_1)\Big]\left(\pp^\mu\pt^\nu+\pp^\nu\pt^\mu\right)\nn\\
&\equiv&\factorglobal f_0(x_1,x_2)f_3^{\mu\nu}(x_1,x_2).
\eea
Finally, the trace in the term $t_9^{\mu\nu}$ is given by
\bea
\Tr\left\{\gn\ks_\perp\gm(\ks_\perp-\ps_\perp)\right\}=4\left[\left(\kt\cdot\pt+\kt^2\right)\gmn+2\kt^\mu\kt^\nu-\left(\pt^\mu\kt^\nu+\pt^\nu\kt^\mu\right)\right].
\eea
After introducing the Schwinger parametrization and performing the sum together with the shift in Eq.~(\ref{qtshift}), we get
\bea
t_9^{\mu\nu}&=&-\frac{2}{\dpi^4}g^2\int\frac{d^2x}{\cosh^2(\eB x_1)\cosh^2(\eB x_2)}\exp\left[-\frac{\tanh(\eB x_1)\tanh(\eB x_2)}{\tanh(\eB x_1)+\tanh(\eB x_2)}\frac{\pt^2}{\eB}\right]\nn\\
&\times&\int d^2\kp e^{\alpha(\kp)x_1}e^{\beta(\kp)x_2}\int d^2\qt e^{-\eta\qt^2}\Big[\left(\qt^2+\sigma(\sigma-1)\pt^2\right)g^{\mu\nu}+2\qt^\mu\qt^\nu+2\sigma(\sigma-1)\pt^\nu\pt^\mu\Big],
\eea
where we have ignored linear terms in $\qt$ and the variables $\eta$ and $\sigma$ are defined in Eqs.~(\ref{etaandsigma}). In Euclidean space, by means of the change of variable given in Eq.~(\ref{ldef}), the parallel integral is easily performed, yielding
\bea
t_9^{\mu\nu}&=&-\frac{2i\pi}{\dpi^4}g^2
\int\frac{d^2x}{(x_1+x_2)\cosh^2(\eB x_1)\cosh^2(\eB x_2)}\exp\left[\frac{x_1x_2}{x_1+x_2}\pp^2-m_f^2(x_1+x_2)\right]\nn\\
&\times&\exp\left[-\frac{\tanh(\eB x_1)\tanh(\eB x_2)}{\tanh(\eB x_1)+\tanh(\eB x_2)}\frac{\pt^2}{\eB}\right]\mathcal{J}_2^{\mu\nu}(x_1,x_2),
\eea
where
\bea
\mathcal{J}_2^{\mu\nu}&=&\int d^2\qt e^{-\eta\qt^2}\Big[\left(\qt^2+\sigma(\sigma-1)\pt^2\right)g^{\mu\nu}+2\qt^\mu\qt^\nu+2\sigma(\sigma-1)\pt^\nu\pt^\mu\Big].
\eea
The last integral has a simple Gaussian form and it is straightforward to compute it, yielding
\bea
\mathcal{J}_2^{\mu\nu}&=&\frac{\pi\eB^2}{\left[\tanh(\eB x_1)+\tanh(\eB x_2)\right]^2}\left[\left(1-\frac{\tanh(\eB x_1)\tanh(\eB x_2)\pt^2}{\eB\left[\tanh(\eB x_1)+\tanh(\eB x_2)\right]}\right)\gmn\right.\nn\\
&-&\left.\gmn_\perp-\frac{2\tanh(\eB x_1)\tanh(\eB x_2)}{\eB\left[\tanh(\eB x_1)+\tanh(\eB x_2)\right]}\pt^\nu\pt^\mu\right].
\eea
Putting all of this together, we get
\bea
t_9^{\mu\nu}&=&-\frac{i\eB^2}{8\pi^2}\int\frac{d^2x}{(x_1+x_2)\sinh^2\left[ºeB(x_1+x_2)\right]}\left[\left(1-\frac{\tanh(\eB x_1)\tanh(\eB x_2)\pt^2}{\eB\left[\tanh(\eB x_1)+\tanh(\eB x_2)\right]}\right)\gmn\right.\nn\\
&-&\left.\gmn_\perp-\frac{2\tanh(\eB x_1)\tanh(\eB x_2)}{\eB\left[\tanh(\eB x_1)+\tanh(\eB x_2)\right]}\pt^\nu\pt^\mu\right]\nn\\
&\equiv&\factorglobal f_0(x_1,x_2)f_4^{\mu\nu}(x_1,x_2).
\eea

\section{Tensor manipulation of Eqs.~(\ref{fs})}\label{ApB}
In order to bring to light the tensor structure of Eq.~(\ref{Pienbaseortonormal}), the terms $f_1^{\mu\nu}(x_1,x_2)$, $f_3^{\mu\nu}(x_1,x_2)$ and $f_4^{\mu\nu}(x_1,x_2)$ in Eqs.~(\ref{fs}) have been factorized in a convenient way, so as to avoid the projection procedure wich can lead non-physical contributions. The tensor $f_2^{\mu\nu}(x_1,x_2)$ remains unchanged and the  manipulation is made by direct inspection.\\

For $f_1^{\mu\nu}(x_1,x_2)$:
\bea
f_1^{\mu\nu}(x_1,x_2)&=&\eB\coth\left[\eB(x_1+x_2)\right]\left[\left(\frac{x_1x_2}{(x_1+x_2)^3}\pp^2+\frac{m_f^2}{x_1+x_2}\right)\gmn_\parallel-\frac{2x_1x_2}{(x_1+x_2)^3}\pmu_\parallel\pnu_\parallel\right]\nn\\
&=&\eB\coth\left[\eB(x_1+x_2)\right]\left[\left(\frac{x_1x_2}{(x_1+x_2)^3}\pp^2+\frac{m_f^2}{x_1+x_2}\right)\gmn_\parallel+\frac{2x_1x_2}{(x_1+x_2)^3}\left(\pp^2\gmn_\parallel-\pp^2\gmn_\parallel-\pmu_\parallel\pnu_\parallel\right)\right]\nn\\
&=&\eB\frac{\coth\left[\eB(x_1+x_2)\right]}{(x_1+x_2)^3}\left[2x_1x_2\pp^2\,\Pp+\left(m_f^2(x_1+x_2)^2-x_1x_2\pp^2\right)\gmn_\parallel\right].
\eea

For $f_3^{\mu\nu}(x_1,x_2)$:

\bea
f_3^{\mu\nu}(x_1,x_2)=\frac{\eB}{2(x_1+x_2)^2\sinh^2\left[\eB (x_1+x_2)\right]}\Big[x_1\sinh(2\eB x_2)+x_2\sinh(2\eB x_1)\Big]\left(\pmu_\parallel\pnu_\perp+\pnu_\parallel\pmu_\perp\right).
\eea
Notice that
\bea
\pmu\pnu=\left(\pp^\mu-\pt^\mu\right)\left(\pp^\nu-\pt^\nu\right)=\pp^\mu\pp^\nu+\pt^\mu\pt^\nu-\left(\pp^\mu\pt^\nu+\pp^\nu\pt^\mu\right),
\eea
therefore,
\bea
\left(\pp^\mu\pt^\nu+\pp^\nu\pt^\mu\right)&=&\pp^\mu\pp^\nu+\pt^\mu\pt^\nu-\pmu\pnu\nn\\
&=&\pp^\mu\pp^\nu+\pt^\mu\pt^\nu-\pmu\pnu+p^2\gmn-p^2\gmn\nn\\
&=&p^2\left(\gmn-\frac{\pmu\pnu}{p^2}\right)+\pp^\mu\pp^\nu+\pt^\mu\pt^\nu-(\pp^2-\pt^2)\left(\gmn_\parallel+\gmn_\perp\right)\nn\\
&=&p^2\left(\gmn-\frac{\pmu\pnu}{p^2}\right)-\pp^2\Pp+\pt^2\Pt-\pp^2\gmn_\perp+\pt^2\gmn_\parallel\nn\\
&=&p^2\left(\gmn-\frac{\pmu\pnu}{p^2}-\Pp-\Pt\right)+p^2\Pp+p^2\Pt-\pp\Pp+\pt^2\Pt-\pp^2\gmn_\perp+\pt^2\gmn_\parallel\nn\\
&=&p^2\Pcero-\pt^2\Pp+\pp^2\Pt-\pp^2\gmn_\perp+\pt^2\gmn_\parallel.
\eea
Thus,
\bea
f_3^{\mu\nu}(x_1,x_2)=\frac{\eB\Big[x_1\sinh(2\eB x_2)+x_2\sinh(2\eB x_1)\Big]}{2(x_1+x_2)^2\sinh^2\left[\eB (x_1+x_2)\right]}\left(p^2\Pcero-\pt^2\Pp+\pp^2\Pt-\pp^2\gmn_\perp+\pt^2\gmn_\parallel\right),\nn\\
\eea
Finally, for $f_4^{\mu\nu}(x_1,x_2)$, given that
\bea
\pmu_\perp\pnu_\perp=\pmu_\perp\pnu_\perp+\pt^2\gmn_\perp-\pt^2\gmn_\perp=\pt^2\Pt-\pt^2\gmn_\perp,
\eea
we have
\bea
f_4^{\mu\nu}(x_1,x_2)&=&\frac{\eB^2}{(x_1+x_2)\sinh^2\left[\eB(x_1+x_2)\right]}\Bigg[
\left(1-\frac{\tanh(\eB x_1)\tanh(\eB x_2)}{\eB\left[\tanh(\eB x_1)+\tanh(\eB x_2)\right]}\pt^2\right)\gmn\nn\\
&-&\gmn_\perp-\frac{2\tanh(\eB x_1)\tanh(\eB x_2)}{\eB\left[\tanh(\eB x_1)+\tanh(\eB x_2)\right]}\pmu_\perp\pnu_\perp\Bigg]\nn\\
&=&\frac{\eB^2}{(x_1+x_2)\sinh^2\left[\eB(x_1+x_2)\right]}\Bigg[
\left(1-\frac{\tanh(\eB x_1)\tanh(\eB x_2)}{\eB\left[\tanh(\eB x_1)+\tanh(\eB x_2)\right]}\pt^2\right)\gmn\nn\\
&-&\gmn_\perp-\frac{2\tanh(\eB x_1)\tanh(\eB x_2)}{\eB\left[\tanh(\eB x_1)+\tanh(\eB x_2)\right]}\Pt+\frac{2\tanh(\eB x_1)\tanh(\eB x_2)}{\eB\left[\tanh(\eB x_1)+\tanh(\eB x_2)\right]}\pt^2\gmn_\perp\Bigg].\nn\\
\eea

By collecting the common terms of the structures $\Pp, \Pt$ and $\Pcero$, we find the coefficients of Eqs.~(\ref{Pipara})-(\ref{coefB}).
\section{Elimination of spurious tensors}\label{ApC}
In order to eliminate the spurious contributions, we follow the procedure discussed in Ref.~\cite{Bjorken}. First, let us scale the $x$ parameters, such that $x_i\rightarrow\lambda z_i$, with $(\lambda, \,z_i)\in \mathbb{R}$. Therefore, the integral that involves the coefficient $A_1$ is

\bea
\mathcal{I}_{A_1}&=&\lambda^2\int d^2z\exp\left[\lambda\left(\frac{z_1 z_2}{z_1+z_2}\pp^2-m_f^2(z_1+z_2)\right)\right]\exp\left[-\frac{\tanh(\lambda\eB z_1)\tanh(\lambda\eB z_2)}{\tanh(\lambda\eB z_1)+\tanh(\lambda\eB z_2)}\frac{\pt^2}{\eB}\right]\nn\\
&\times&\Bigg\{\frac{\coth\left[\lambda\eB(z_1+z_2)\right]}{\lambda(z_1+z_2)^3}\left(m_f^2(z_1+z_2)^2-z_1z_2\pp^2\right)+\frac{z_1\sinh(2\eB z_2)+z_2\sinh(2\lambda\eB z_1)}{2\lambda(z_1+z_2)^2\sinh^2\left[\lambda\eB (z_1+z_2)\right]}\pt^2\nn\\
&+&\frac{\eB}{\lambda(z_1+z_2)\sinh^2\left[\lambda\eB(z_1+z_2)\right]}\left(1-\frac{\tanh(\lambda\eB z_1)\tanh(\lambda\eB z_2)}{\eB\left[\tanh(\lambda\eB z_1)+\tanh(\lambda\eB z_2)\right]}\pt^2\right)\Bigg\},
\eea
which can be written as
\bea
\mathcal{I}_{A_1}&=&-\lambda\frac{\partial}{\partial\lambda}\int\frac{d^2z}{(z_1+z_2)^2}\coth\left[\lambda\eB(z_1+z_2)\right]\exp\left[\lambda\left(\frac{z_1 z_2}{z_1+z_2}\pp^2-m_f^2(z_1+z_2)\right)\right]\nn\\
&\times&\exp\left[-\frac{\tanh(\lambda\eB z_1)\tanh(\lambda\eB z_2)}{\tanh(\lambda\eB z_1)+\tanh(\lambda\eB z_2)}\frac{\pt^2}{\eB}\right].
\eea
Scaling back $\lambda\,z_1\rightarrow x_i$, we obtain
\bea
\mathcal{I}_{A_1}&=&-\lambda\frac{\partial}{\partial\lambda}\int\frac{d^2x}{(x_1+x_2)^2}\coth\left[\eB(x_1+x_2)\right]\exp\left[\frac{x_1 x_2}{x_1+x_2}\pp^2-m_f^2(x_1+x_2)\right]\nn\\
&\times&\exp\left[-\frac{\tanh(\eB x_1)\tanh(\eB x_2)}{\tanh(\eB x_1)+\tanh(\eB x_2)}\frac{\pt^2}{\eB}\right],
\eea
and thus, the derivative is applied to a function independent of $\lambda$. Therefore $\mathcal{I}_{A_1}=0$.

The implementation of the same argument for $\mathcal{I}_{A_2}$ is more involved, given that the function is not a trivial combination of coefficients for $\pp^2$ and $\pt^2$. After the $\lambda$-scaling, the integral is
\bea
\mathcal{I}_{A_2}=\lambda^2\int\frac{d^2z}{\lambda(z_1+z_2)^2}I(\lambda z_1,\lambda_2),
\eea
where
\bea
I(\lambda z_1,\lambda z_2)&=&f_0(\lambda z_1,\lambda z_2) \Bigg{[}\frac{\cosh\left[\lambda\eB(z_2-z_1)\right]}{\sinh\left[\lambda\eB(z_1+z_2)\right]}\left(\frac{z_1z_2}{z_1+z_2}\pp^2+m_f^2(z_1+z_2)+\frac{1}{\lambda}\right)\nn\\
&-&\frac{z_1\sinh(2\lambda\eB z_2)+z_2\sinh(2\lambda\eB z_1)}{2\sinh^2\left[\lambda\eB (z_1+z_2)\right]}\pp^2+\frac{(z_1+z_2)\sinh(\lambda\eB z_1)\sinh(\lambda\eB z_2)}{\sinh^3\left[\lambda\eB(z_1+z_2)\right]}\pt^2\Bigg{]},
\eea
so that by expanding in a Taylor series around $\lambda=0$ it is possible to find that

\bea
\int I(\lambda z_1,\lambda z_2) d\lambda&=&-\frac{1}{\eB(z_1+z_2)\lambda}+\frac{2\eB^2(z_1+z_2)^2(z_1^2-4z_1z_2+z_2^2)+3\left(z_1z_2p^2-m^2(z_1+z_2)^2\right)^2}{6\eB (z_1+z_2)^3}\lambda\nn\\
&+&\frac{\lambda^2}{6\eB (z_1+z_2)^4}\Bigg{[}\left(3p^4z_1^2z_2^2-3m^2p^2z_1z_2(z_1+z_2)^2+m^4(z_1+z_2)^4\right)(z_1+z_2)^2m^2\nn\\
&-&z_1z_2\left(p^6z_1^2z_2^2-2\eB^2(z_1+z_2)^2\left(p^2(z_1-z_2)^2-\pt^2z_1z_2\right)\right)\Bigg{]}\nn\\
&+&\frac{\lambda^3}{1080\eB(z_1+z_2)^5}\Bigg{[}45\left(p^2z_1z_2-m^2(z_1+z_2)^2\right)^4\nn\\
&+&8\eB^4(z_1+z_2)^4\left(z_1^4+4z_1^3z_2-24z_1^2z_2^2+4z_1z_2^3+z_2^4\right)\nn\\
&-&60\eB^2(z_1+z_2)^2\left(p^2z_1z_2-m^2(z_1+z_2)^2\right)\left(m^2(z_1+z_2)^2(z_1^2-4z_1z_2+z_2^2)\right.\nn\\
&+&\left.z_1z_2\left(p^2(3z_1^2-4z_1z_2+3z_2^2)-6\pt^2z_1z_2\right)\right)\Bigg{]}+\mathcal{O}\left(\lambda^4\right),
\eea

where the desired scaling properties are recovered and hold for all orders in $\lambda$. This means that is possible to write
\bea
\int I(\lambda z_1,\lambda z_2) d\lambda&=&-\frac{1}{\eB(z_1+z_2)\lambda}+h\left(\lambda z_1,\lambda z_2\right),\nn\\
\eea
thus
\bea
I(\lambda z_1,\lambda z_2)&=&\frac{\partial}{\partial\lambda}\left[-\frac{1}{\eB(z_1+z_2)\lambda}+h\left(\lambda z_1,\lambda z_2\right)\right]=\frac{\partial}{\partial\lambda}\left[-\frac{1}{\eB(x_1+x_2)}+h\left(x_1, x_2\right)\right]=0,
\eea
and therefore, $\mathcal{I}_{A_2}=0$.

The above argument is valid for all values of $\lambda$. Consequently, the result can be taken as general.

\end{widetext}


\end{document}